\documentclass[runningheads,11pt]{llncs} 
\usepackage[T1]{fontenc}
\usepackage[letterpaper, margin=1.3in]{geometry}
\newcommand{\citet}[1]{\cite{#1}}
\newcommand{\citep}[1]{\cite{#1}}

\usepackage[ruled]{algorithm2e} 

\SetAlFnt{\small}
\SetAlCapFnt{\small}
\SetAlCapNameFnt{\small}
\SetAlCapHSkip{0pt}
\IncMargin{-\parindent}
\usepackage{lineno}

\usepackage[utf8]{inputenc}

\usepackage[ruled]{algorithm2e} 
\usepackage{algorithmic} 
\usepackage{pifont}
\usepackage{nicefrac}
\usepackage{color}
\usepackage{xcolor,colortbl}
\usepackage{wrapfig}
\usepackage{graphicx}
\usepackage{amsmath,amsthm}
\usepackage{amssymb}
\usepackage{mathtools}
\usepackage{arydshln}
\usepackage{diagbox}
\usepackage[a]{esvect}
 
\usepackage{bm}
\usepackage{rotating}
\usepackage{multicol}
\usepackage{multirow}
\usepackage{latexsym}
\usepackage{enumitem}
\usepackage{hyperref}[backref=page]
\usepackage{float}
\hypersetup{
     colorlinks   = true,
     linkcolor    = red, 
     urlcolor     = blue, 
	 citecolor    = blue 
} 

\usepackage[framemethod=TikZ]{mdframed}
\newcounter{theo}[section] 

\newtheorem{thm}{Theorem}
\newtheorem{dfn}{Definition}

\newtheorem{lem}{Lemma}
\newtheorem{ex}{Example}

\newtheorem{prop}{Proposition}
\newtheorem{coro}{Corrollary}
\newenvironment{sketch}{\noindent{\bf Proof sketch.}\rm }{\hfill $\Box$ }

\newcounter{newct}

\newcommand{\bv}{\begin{array}}

\newcommand{\appClaim}[3]{\vspace{1mm}   {\sc Claim~\ref{#2}.} {\bf (#1). } {\em #3}}
\newcommand{\appLem}[3]{\vspace{1mm}\noindent{\bf Lemma~\ref{#2}.} {\bf (#1). } {\em #3}}
\newcommand{\appThm}[3]{\vspace{1mm} {\sc Theorem~\ref{#2}} {({\bf #1}). }{\em #3 \vspace{1mm}}}

\newcommand{\appThmnoname}[2]{\vspace{1mm} {\sc Theorem~\ref{#1}.} {\em #2 \vspace{1mm}}}

\newcommand{\myparagraph}[1]{\vspace{2mm}\noindent {\bf\boldmath #1}}

\renewcommand{\cal}{\mathcal }
\newcommand{\ma}{\mathcal A}

\newcommand{\calP}{\mathcal P}

\newcommand{\ml}{\mathcal L}

\newcommand{\calS}{\mathcal S}

\newcommand{\ra}{\rightarrow}

\newcommand{\hpp}{\text{MFP}}
\newcommand{\hp}{\text{MFP}}
\newcommand{\cor}{{\overline r}} 

\newcommand{\rank}{\text{Rank}}

\newcommand{\tb}{f}

\newcommand{\stab}{\text{\sc Stab}}
\newcommand{\orbit}{\text{\sc Orbit}}
\newcommand{\stabset}[2]{\text{\sc Stab}_{#1}^{#2}}
\newcommand{\orbitset}[2]{\text{\sc Orbit}_{#1}^{#2}}
\newcommand{\fixed}{\text{\sc Fixed}}
\newcommand{\fw}{\text{\sc FPD}}
\newcommand{\hist}{\text{Hist}}
\newcommand{\histset}[3]{{\mathbb H}_{#1,#2}^{#3}}

\newcommand{\id}{\text{Id}}

\newcommand{\Omit}[1]{}

\newcommand{\calC}{\mathcal C}

\newcommand{\calD}{\mathcal D}

\newcommand{\wmg}{\text{WMG}}

\newcommand{\copeland}{\text{Cd}_\alpha}

\newcommand{\veto}{\text{veto}}

\newcommand{\sgroup}{{\cal S}_{\ma}}

\newcommand{\ano}{\text{\sc Ano}}
\newcommand{\neu}{\text{\sc Neu}}
\newcommand{\res}{\text{\sc Res}}

\newcommand{\anr}{\text{\sc ANR}}

\newcommand{\orbsize}[2]{{\text{\rm Sizes}({#2}/#1)}}

\newcommand{\lcm}{\text{\rm lcm}}
\newcommand{\intcomb}[1]{{\text{\rm Feas$\left(#1\right)$}{}}}

\newcommand{\vpl}[2]{{{\cal E\hspace{-1pt}\cal L}_{#2}(#1)}}
\newcommand{\vpa}[2]{{{\cal E\hspace{-1pt}\cal M}_{#2}(#1)}}
\newcommand{\vda}[2]{{{\cal D\hspace{-1pt}\cal M}_{#2}(#1)}}
\newcommand{\vdl}[2]{{{\cal D\hspace{-1pt}\cal L}_{#2}(#1)}}

\newcommand{\prefspace}{{\cal E}}
\newcommand{\decspace}{{\cal D}}
\newcommand{\listset}[1]{{\cal L}_{#1}}
\newcommand{\committee}[1]{{\cal M}_{#1}}

\newcommand{\commonsettings}{\text{Common} \text{Settings}} 

\newcommand{\simplex}[2]{{{\calS}_{#1}^{#2}}}
\newcommand{\lcmset}[2]{{\text{\rm Coins}^{#2}(#1)}}
\newcommand{\vecp}[1]{{\vec {#1}\,'}}

\newcommand{\breaking}[2]{(#1\ast #2)}

\newcommand{\problematic}[4]{\calP_{#1}}

\usepackage[most]{tcolorbox}
\newtcolorbox{KRQ}[2][]{left=0pt,right=0pt,boxrule=.4pt,
                lower separated=false,
                colback=white,
colframe=black,fonttitle=\bfseries,
colbacktitle=black!20,
coltitle=black,
enhanced,
attach boxed title to top left={yshift=-0.1in,xshift=0in},
                 boxed title style={boxrule=.4pt,colframe=black,},
title=#2,#1}

\begin{document}
\title{Most Equitable Voting Rules}
\author{
Lirong Xia
}  
\institute{RPI\\xialirong@gmail.com}
\maketitle

\begin{abstract} 
In social choice theory, {\em anonymity} (all agents being treated equally) and {\em neutrality} (all alternatives being treated equally) are widely regarded as ``{\em minimal demands}'' and ``{\em uncontroversial}'' axioms  of equity and fairness. However, the {\em ANR impossibility}---there is no voting rule that satisfies anonymity, neutrality, and resolvability (always choosing one winner)---holds even in the simple setting of two alternatives and two agents.  How to design voting rules that optimally satisfy anonymity, neutrality, and resolvability remains an open question.

We address the optimal design question for a wide range of preferences and decisions that include ranked lists and committees. Our conceptual contribution is a novel and  strong notion of {\em most equitable refinements} that optimally preserves anonymity and neutrality for any irresolute rule that satisfies the two axioms. Our technical contributions are twofold. First, we characterize the conditions for the ANR impossibility to hold under general settings, especially when the number of agents is large. Second, we propose the {\em most-favorable-permutation (MFP)} tie-breaking to compute a most equitable refinement and design a polynomial-time algorithm to compute MFP when agents' preferences are full rankings.
 \keywords{Social choice, anonymity, neutrality, tie-breaking}
\end{abstract}

\section{Introduction}

A major goal of social choice is to design a {resolute} voting rule  $r:\prefspace^n\ra \decspace$ that  maps $n$ agents' preferences, each of which is chosen from the preference space  $\prefspace$, to a single collective decision in the decision space $\decspace$.    For example, in single-winner elections, each agent uses a linear order over the  alternatives  to represent his/her preferences, and the decision is a single alternative, i.e., the winner.



So how can we design the best voting rule? The answer depends on the measure of goodness. In axiomatic social choice, various normative measures of voting, called {\em axioms}, were proposed to evaluate and design voting rules~\citep{Plott76:Axiomatic}.  While different  axioms  are desirable in different scenarios, the two equity/fairness axioms known as {\em anonymity} (all agents being treated equally)  and {\em neutrality} (all alternatives being treated equally) are broadly viewed as ``{\em minimal demands}'' and ``{\em uncontroversial}''~\citep{Zwicker2016:Introduction,Myerson2013:Fundamentals,Brandt2019:Collective}. 

Indeed, it is easy to satisfy anonymity and neutrality if we allow ties by using an {\em irresolute} rule $\cor:\prefspace^n\ra (2^{\decspace}\setminus \emptyset)$. For example, both axioms are satisfied by the irresolute rule that always chooses all decisions $\decspace$ regardless of agents' preferences, and by the majority rule with ties when there are two alternatives~\citep{May52:Set}. However, ties are not allowed in many scenarios, and in such cases a resolute rule must be used.

Consequently, it is natural and desirable to design voting rules that simultaneously satisfy anonymity, neutrality, and {\em resolvability} (i.e., the rule must be resolute). Such rules are called {\em ANR  rules}~\citep{Dogan2022:Anonymous}. Unfortunately, no ANR rule exists even  under the single-winner election setting with two alternatives and two agents, as shown in the following simple proof.

\begin{KRQ}{Proof: ANR impossibility}
\vspace{1mm}
Let $\{1,2\}$ denote the two alternatives. Consider a ``problematic'' preference profile $P= (1\succ 2, 2\succ 1)$  and a permutation $\sigma$ that exchanges the names of the two alternatives. Suppose there exists an ANR rule $r$. W.l.o.g.~suppose $r(P) = 1$.  Then, by neutrality $r(\sigma(P)) =  \sigma(r(P)) = 2$ and by anonymity $r(\sigma(P)) = r(P)=1$, which is a contradiction. 
\end{KRQ}

This incompatibility between anonymity, neutrality, and resolvability is ``{\em among the most well-known results in social choice theory}''~\citep{Ozkes2021:Anonymous} and is often presented as a first course when discussing fairness/equity in social choice, see, e.g.,~\citep{Myerson2013:Fundamentals,Zwicker2016:Introduction,Lackner2023:Multi-Winner}. Perhaps because it is so fundament and the proof is so simple, it is often viewed as a folklore  and doesn't have a name of its own. Following the convention~\citep{Dogan2011:Anonymous,Dogan2015:Anonymous,Dogan2022:Anonymous}, we call it the {\em ANR impossibility} in this paper. In fact, Moulin~\citet{Moulin1983:The-Strategy} proved that 
the ANR impossibility holds  if and only if  the number of alternatives $m$ can be represented as the sum of the number of agents $n$'s non-trivial (i.e., $>1$) divisors.   

Surprisingly, despite the significance and desirability of anonymity and neutrality, our understanding of how to achieve them is still  limited, and little is known beyond Moulin's characterization~\citet{Moulin1983:The-Strategy}. Specifically, when ANR rules exist for certain $m$ and $n$, it is unclear if any of them is polynomial-time computable. When ANR rules do not exist, it is unclear what is an informative measure of equity w.r.t.~anonymity and neutrality. One natural approach is to consider the likelihood of  ANR violations when agents' preferences are generated from some statistical model~\citep{Xia2020:The-Smoothed}, yet how to choose an appropriate model  and how to design rules with minimal likelihood of ANR violations still remains open.

Therefore, taking anonymity and neutrality as fundamental notions of equity, it is natural to ask how to {\em optimally} achieve equity in voting, i.e.: 
\begin{center}
{\bf How can we design most equitable voting rules?} 
\end{center}
The same question also arises in other common social choice settings with various combinations of preference space and decision space. For example, in the Arrovian framework~\citep{Campbell2002:Impossibility}, an agent ranks all alternatives and the collective decision is a ranking over the alternatives. In rank aggregation~\citep{Dwork01:Rank}, an agent ranks a subset of $\ell$ alternatives and the collective decision is a ranking over $k$ alternatives. In approval voting~\citep{Fishburn1978:Approval}, each agent ``approves'' a set of  $\ell$ alternatives and the collective decision is a single alternative. In multi-winner elections~\citep{Lu2011:Budgeted,Elkind2017:Properties,Lackner2023:Multi-Winner}, an agent's preferences are represented by a ranking or a set of approved alternatives, and the collective decision is a set of $k$-committee that consists of $k$ alternatives. 
As a forth example, in the 2021 New York City Democratic mayoral primary election, an agent can rank up to $5$ alternatives and the collective decision is a single alternative. Among these settings, the condition for  the ANR impossibility to hold was only known when an agent's preferences and the decision are both linear orders over all alternatives, due to a characterization by Bubboloni and Gori~\citep{Bubboloni2014:Anonymous}: the ANR impossibility holds if and only if $m\ge n$'s smallest non-trivial divisor. 


\subsection{Our Contributions}
\label{sec:contributions}
We investigate  general preference space $\prefspace$ and decision space $\decspace$, with a focus on the {\em Common Settings} of $(\prefspace,\decspace)$ defined as follows, where the preferences and decisions are ranked lists or committees.

\begin{tcolorbox}The {\em\bf \commonsettings{}} in this paper refer to  
$$(\prefspace,\decspace)\in \{\listset \ell, \committee \ell: 1\le \ell\le m\}\times \{\listset k, \committee k: 1\le k\le m\},$$
where $m$ is the number of alternatives; for any $i\le m$, $\listset i$ is the set of all ranked lists over $i$ alternatives  and  $\listset i$ is the set of all  $i$-committees (subsets of $i$ alternatives).
\end{tcolorbox}

The Common Settings cover many common social choice settings discussed above, as shown in the following table.
\begin{center}
\renewcommand{\arraystretch}{1}
\begin{tabular}{|c|*{2}{c|}}
 \hline
 \diagbox{$\prefspace$}{$\calD$} & $\listset k$ & $\committee k$ \\
 \hline
 \rule{0pt}{15pt}$\listset \ell$ & \begin{tabular}{@{}c@{}}single-winner elections, \\Arrovian framework~\citep{Campbell2002:Impossibility}\\rank aggregation~\citep{Dwork01:Rank}\end{tabular} & multi-winner voting~\citep{Elkind2017:Properties}  \\[5pt]
\hline
\rule{0pt}{15pt}$\committee \ell$ &  approval voting~\citep{Fishburn1978:Approval} & approval-based committee voting~\citep{Lackner2023:Multi-Winner} \\[5pt]
\hline
 \end{tabular}
 \renewcommand{\arraystretch}{1}
\end{center} 

\myparagraph{Overall approach.} We approach the optimal design problem  via tie-breaking. More precisely, for  any  irresolute  rule $\cor$ that satisfies anonymity and neutrality, we aims at designing a {\em refinement}, which is a resolute voting rule $r$ that chooses a single co-winner of $\cor$ as the sole winner, to optimally satisfy anonymity and neutrality (while resolvability is automatically satisfied). This approach is not only a natural common practice~\citep{Zwicker2016:Introduction}, but also without loss of generality, because any voting rule can be viewed as applying a tie-breaking mechanism to the irresolute rule $\cor_\decspace$ that always chooses all decisions.

\myparagraph{\bf Our conceptual contribution} is a novel notion of {\em most equitable refinements} for any  irresolute  rule $\cor$ that satisfies anonymity and neutrality, especially $\cor_\decspace$. This is a strong notion of optimality and is not guaranteed to exist by definition, because it requires that a most equitable refinement of $\cor$ achieves the same or higher ANR satisfaction than {\em every} refinement of $\cor$ at {\em every} preference profile. Consequently, a most equitable refinement of $\cor$ also has the same or higher probability to satisfy ANR  than {\em every} refinement of $\cor$ under {\em every} distribution over agents' preferences. Surprisingly, most equitable refinements always exists (Lemma~\ref{lem:anr-characterization}). We are not aware of a similar notion in the literature.


\myparagraph{\bf Our technical contributions} are two-fold. 

\myparagraph{\bf  First: Characterizations of the ANR impossibility.} It follows from the optimality of most equitable refinements that the ANR impossibility holds if and only if any most equitable refinement of $\cor_\decspace$ satisfies anonymity and neutrality. Leveraging this observation, we characterize  conditions on $m$ and $n$ for the ANR impossibility to hold under the \commonsettings{}   (Theorem~\ref{thm:ANR-common}): it holds if and only if a {\em partition condition} is met, i.e., there exists an integer partition $\vec m$ of $m$ that satisfies a {\em sub-vector constraint}, and a {\em change-making constraint}, which requires that $n$ can be made up by coins whose denominations depend on $m$. This characterization not only resolves the open question on the ANR impossibility for \commonsettings{}, but also provides a novel angle that unifies existing characterizations for  $(\prefspace, \decspace) = (\listset m, \listset 1)$~\citep{Moulin1983:The-Strategy} and $(\listset m, \listset m)$~\citep{Bubboloni2014:Anonymous}. 
 We also apply Theorem~\ref{thm:ANR-common}  to 
characterize {\em at-large ANR impossibility}  (Theorem~\ref{thm:at-large-up-to-L}), i.e., the ANR impossibility holds for every sufficiently large $n$, for up-to-$L$ lists and committees. 
As a corollary, the ANR impossibility holds for the 2021 New York City Democratic mayoral primary elections (Example~\ref{ex:NYC}). 

\myparagraph{\bf Second: Computing most equitable refinements.} We propose the {\em most-favorable-permutation (MFP)} tie-breaking mechanisms to obtain  most equitable refinements, and design a polynomial-time algorithm (Algorithm~\ref{alg:tie-breaking}) to compute them when $\prefspace = \listset m$.  A straightforward application of MFP tie-breaking to $\cor_\decspace$ gives us a polynomial-time ANR rule when Moulin~\citet{Moulin1983:The-Strategy}'s condition or Bubboloni and Gori~Bubboloni and Gori~\citet{Bubboloni2016:Resolute}'s condition is not satisfied, thus addressing computational challenges identified in previous work~\citep{Bubboloni2021:Breaking}.

\myparagraph{Technical innovations.} Our work builds upon notation and principles of {\em algebraic voting theory}~\citep{Daugherty2009:Voting} and the group theoretic framework for analyzing anonymity and neutrality~\citep{Egecioglu2013:The-Impartial,Dogan2022:Anonymous}. While most previous work focused on using the framework to characterize the conditions for the ANR impossibility to hold, we take a step further by developing the framework to  investigate  optimal refinements of irresolute rules for general preferences and  decisions. As discussed above, our framework can naturally be used to obtain and extend previous characterizations of the ANR impossibility, as shown in Theorem~\ref{thm:ANR-common}.  
The key technical innovations in the proof of Theorem~\ref{thm:ANR-common} are novel applications of the orbits-stabilizer theorem in group theory. 
Our Algorithm~\ref{alg:tie-breaking}  addresses the computational challenge pointed out in previous work~\citep{Bubboloni2021:Breaking}  by exploring a simple and efficient way to identify  ``representative'' profiles. 
While the main text of the paper focuses on the \commonsettings{}, our  methodology naturally generalizes to even more general settings as discussed in Appendix~\ref{sec:general}.

 \section{Preliminaries}
\label{sec:prelim}

\noindent{\bf Decisions.} Let $\ma=[m] = \{1,\ldots, m\}$ denote the set of $m\ge 2$ {\em alternatives}. For any $1\le k\le m $, let  $\committee k$ denote the set of all {\em $k$-committees} of $\ma$, which are sets of $k$ alternatives in $\ma$. That is, $\committee k\triangleq \{A\subseteq \ma: |A| = k\}$. Let $\listset k$ denote the set of all {\em $k$-lists}, each of which is a linear order over $k$ alternatives. That is, $\listset k\triangleq \{\ml(A): A\in \committee k\}$, where $\ml(A)$ is the set of all linear orders over $A$. Let $\calD$ denote the decision space.  Common choices of $\calD$ include $\committee k$ and $\listset k$ for some $k\le m$. 

\myparagraph{\bf Preferences.} There are  $n\in\mathbb N$ agents, each of which uses an element in the {\em preference space} $\prefspace$ to represent his or her preferences, called a {\em vote}. Common choices of $\calD$ include  $\committee \ell$ and $\listset \ell$ for some $1\le\ell\le m$. For example, when $m=5$, $\{1,5\}\in\committee 2$ and $1\succ 2\succ 3\in \listset 3$, where $\succ$ reads ``preferred to''. We  also consider {\em up-to-$L$ committees} that consist  of all $\ell\le L$ committees, formally defined as $\committee {\le L}\triangleq \bigcup_{\ell\le L} \committee \ell$. The {\em up-to-$L$} list $\listset {\le L}$  is defined similarly.  The vector of $n$ agents' votes, denoted by $P$, is called a {\em (preference) profile}, sometimes called an $n$-profile. Given $\prefspace$, let $\hist(P)\in {\mathbb Z}_{\ge 0}^{|\prefspace|}$ denote the {\em histogram} of $P$, which is the anonymized $P$ that contains the total number of times  each element in $\prefspace$ appears in  $P$.   Let $\histset{m}{n}{\prefspace}$ denote the set of all histograms of  $n$-profiles over $\prefspace$. 

\myparagraph{\bf Voting rules.} An {\em  irresolute}   voting rule $\cor: \prefspace^n\ra(2^\decspace\setminus\emptyset)$ maps  a profile to a non-empty subset of $\decspace$. A {\em resolute} voting rule $r: \prefspace^n\ra \decspace$ can be viewed as an irresolute rule that always chooses a single decision. In this paper, we use $\bar{\cdot}$ to indicate irresolute rules, meaning that it is possible, though not guaranteed, that there are two or more winners. We slightly abuse the notation by using $r(P)=d$ and $r(P)= \{d\}$ interchangeably. A voting rule $\cor'$ is a {\em refinement} of another voting rule $\cor$ if for all profiles $P$, $\cor'(P)\subseteq \cor(P)$. 




\begin{ex}[{\bf Positional scoring rules}{}]
When $(\prefspace,\decspace)=(\listset m,\listset 1)$, an irresolute  {\em positional scoring rule} $\cor_{\vec s}$  is characterized by a scoring vector $\vec s=(s_1,\ldots,s_m)$ with $s_1\ge s_2\ge \cdots\ge s_m$ and $s_1>s_m$. For any alternative $a$ and any linear order $R\in\ml(\ma)$, we let $\vec s(R,a)\triangleq s_i$, where $i$ is the rank of $a$ in $R$. Given a profile $P$, let $\vec s(R,a)\triangleq \sum_{R\in P} \vec s(R,a)$ denote the {\em total score} of $a$. When  $(\prefspace,\decspace)=(\listset m,\committee k)$, $\cor_{\vec s}$ chooses the set of all  $k$-committees whose scores are not smaller than that of any other alternative. 
 Special positional scoring rules include {\em plurality}, whose scoring vector is $(1,0,\ldots,0)$,   {\em Borda}, whose scoring vector is $(m-1,m-2,\ldots,0)$, and {\em veto}, whose scoring vector is $(1, \ldots,1,0)$. 
 \end{ex}

\myparagraph{\bf Tie-breaking mechanisms.}  Many commonly studied resolute voting rules are defined as the result of a {\em tie-breaking mechanism}   applied to the outcome of an irresolute rule. A tie-breaking mechanism $\tb$ is a mapping from a profile $P$ and a non-empty set $D\subseteq \calD$ to a single decision in $D$. That is, $f:\prefspace^n\times (2^\decspace\setminus \emptyset)\ra \decspace$.  For example, when $\calD = \ma$, the {\em agenda tie-breaking} breaks ties in favor of alternatives ranked higher w.r.t.~a pre-defined ranking; and the {\em fixed-voter tie-breaking} breaks ties using a pre-defined agent's  ranking.  
Let $\tb\ast\cor$ denote the refinement of $\cor$ by applying $f$. That is, for any profile $P$, $(\tb\ast\cor)(P) = \{\tb(P,\cor(P))\}$. 

\myparagraph{Permutations.} Let $\sgroup$ denote the set of all permutations over $\ma$, called the {\em permutation group}. A permutation can be represented by its cycle form. For example, when $m=4$, 
$(1,3) (2,4)$ represents the permutation that exchanges $1$ and $3$, and also exchanges $2$ and $4$.  Any permutation $\sigma\in \sgroup$ can be naturally extended to  $k$-committees, $k$-lists, profiles, and histograms over $\ma$ as follows. For any $k$-committee $M = \{a_1,\ldots,a_k\}$, let $\sigma(M)\triangleq \{\sigma(a_1),\ldots,\sigma(a_k)\}$; for any $k$-list $R = [a_1\succ \cdots \succ a_k]$, let $\sigma(R)\triangleq [\sigma(a_1)\succ \cdots\succ\sigma(a_k)]$; for any profile $P = (R_1,\ldots, R_n)$, let $\sigma(P) \triangleq (\sigma(R_1),\ldots, \sigma(R_n))$; 
and for any histogram $\vec h \in \histset{m}{n}{\prefspace}$, let $\sigma(\vec h)$ be the histogram such that for every ranking $R$, $[\sigma(\vec h)]_R = [\vec h]_{\sigma^{-1}(R)}$, where $[\sigma(\vec h)]_R$ is the value of the $R$-component in $\sigma(\vec h)$, i.e., the multiplicity of $R$-votes in $\sigma(\vec h)$.

\begin{ex}
\label{ex:perm}
Let $m=5$, $n=10$,  and $P^{+}$ denote any $10$-profile with the following histogram. 
\begin{center}
  \begin{tabular}{|r|c|c|c|c|c|c|}
\hline Ranking& $13245$ & $23145$ & $34125$ & $34215$ & $43125$ & $43215$ \\
\hline \#  in $\hist(P^{+} )$ & $2$&$2$& $2$& $2$& $1$&$1$  \\
\hline
\end{tabular}
\end{center}
In the table, $13245$ represents the ranking $1\succ 3\succ 2\succ 4\succ 5$.  
Let $\sigma_1 = (1,3)(2,4)$, which maps $34125$ to $12345$, and let $\sigma_2 = (1,3,2,4)$, which maps $43125$ to $12345$.  Then, $ \sigma_1(\hist(P^{+}))  $  and $ \sigma_2(\hist(P^{+}))  $ are:
\begin{center}
$ $\begin{tabular}{|r|c|c|c|c|c|c|c|c|}
\hline Ranking& $12345$ & $12435$ & $21345$ & $21435$ & $31425$& $32415$ & $41325$& $42315$ \\
\hline  \#  in $\hist(\sigma_1(P^{+}))$& $2$&$2$& $2$& $2$& $1$ & $0$&$1$ & $0$\\
\hline  \#  in $\hist(\sigma_2(P^{+}))$& $2$&$2$& $2$& $2$& $0$ & $1$&$0$ & $1$\\
\hline
\end{tabular}
\end{center}
\end{ex}

\myparagraph{\bf Anonymity, neutrality, and resolvability.} 
For any rule $\cor$ and any profile $P$, we define $\ano(\cor,P) \triangleq  1$ if  for any profile $P'$ with $\hist(P')=\hist(P)$,   $\cor(P')=\cor(P)$; otherwise $\ano(\cor,P) \triangleq  0$. We define $\neu(\cor,P) \triangleq  1$ if   for every permutation $\sigma$ over $\ma$, we have $\cor(\sigma(P))=\sigma(\cor(P))$; otherwise $\neu(\cor,P) \triangleq 0$. We define $\res(\cor,P) \triangleq  1$ if $|\cor(P)|=1$; otherwise $\res(\cor,P) \triangleq  0$. Notice that a resolute rule outputs a single {\em decision}, which may not be a single alternative---for example when $\decspace = \committee 2$, a decision is a set of two alternatives. If $\ano(\cor,P) = 1$ (respectively, $\neu(\cor,P) = 1$ or $\res(\cor,P) = 1$), then we say that $\cor$ satisfies anonymity (respectively, neutrality or resolvability) at $P$. We further define $\anr(\cor,P)  \triangleq \ano(\cor,P)\times \neu(\cor,P)\times \res(\cor,P)$. That is, $\anr(\cor,P)=1$ ($\cor$ satisfies {\em ANR} at $P$) if and only if $\cor$ satisfies anonymity, neutrality, and resolvability at $P$. Given $n$, we say that $\cor$ satisfies anonymity (respectively, neutrality, resolvability, or ANR) if and only if for all $n$-profiles $P$, we have $\ano(\cor,P) = 1$ (respectively, $\neu(\cor,P) = 1$, $\res(\cor,P) = 1$, or $\anr(\cor,P)=1$). 

In this paper, the {\em  ANR impossibility} refers to the claim that no ANR rule exists given a certain combination of $(\prefspace,\decspace)$,  $m$, and $n$. 
 
\section{Most Equitable Refinements}
\label{sec:MER}
Before formally presenting the definition, let us first examine the following example of a ``problematic'' profile under veto, in which $\anr$ fails under every refinement of veto.
\begin{ex}[{\bf\boldmath A problematic profile under veto}{}]
\label{ex:problematic-veto}
Let $m=5$, $n=10$, $(\prefspace,\decspace)=(\listset m, \listset 1)$,   and  $P^-$ be an arbitrary $10$-profile whose histogram is: 
\begin{center}
  \begin{tabular}{|r|c|c|c|c|c|c|}
\hline Ranking& $13245$ & $23145$ & $31254$ & $32154$ & $41253$ & $42153$ \\
\hline \#  in $\hist(P^- )$ & $2$&$2$& $2$& $2$& $1$&$1$  \\
\hline
\end{tabular}
\end{center}
We have $ \veto(P^{-}) = \{1,2\}$. Let $\sigma $ denote the permutation that exchanges $1$ and $2$ while keeping all other alternatives the same. 
Suppose there exists a refinement $r$ of $\veto$ that satisfies $\anr$ at $P^{-}$. If $r(P^{-}) = \{1\}$, then by neutrality,  $r(\sigma(P^-)) = \{2\}$. On the other hand, notice that $\hist(\sigma(P^-)) = \hist(P^-)$. Therefore, by anonymity, $r(\sigma(P^-))=r(P^-) = \{1\}$, which is a contradiction. A similar contradiction happens if  $r(P^{-})=\{2\}$. Therefore, ANR is guaranteed to fail at $P^-$ under every refinement of veto.
\end{ex}
Other problematic profiles exist under veto when $m=5$ and $n=10$. Therefore, the best we could possibly achieve in a refinement is to preserve $\anr$ at all non-problematic profiles. And if such a refinement exists, then we call it a {\em most equitable refinement}, formally defined as follows. 
\begin{dfn}[{\bf Problematic profiles and most equitable refinements}{}]
For any irresolute  rule $\cor$ and any  $n\ge 1$, let $\problematic{\cor}{n}{\prefspace}{\decspace}$ denote the set of {\em problematic profiles} $P$, such that 
for every refinements $r$ of $\cor$, $\anr(r, P)=0$. A refinement $r^*$ of $\cor$ is called a {\em most equitable refinement}, if $\anr(r^*, P)=1$ for every $P\notin \problematic{\cor}{n}{\prefspace}{\decspace}$.
\end{dfn}

A most equitable refinement may not exist by definition, because  a refinement may satisfy $\anr$ at one non-problematic profile but not at another. Therefore, most equitable refinements are a strong notion of optimality. Put in another way, a most equitable refinement has the same  or higher likelihood of $\anr$ satisfaction than any other refinement w.r.t.~any distribution over the profile. It is surprising to see that they indeed exist according to the following lemma. 

\begin{lem}[{\bf Existence of most equitable refinements}{}]
\label{lem:anr-characterization}
Under \commonsettings{}, any anonymous and neutral  rule  has a most equitable refinement.  
\end{lem}

\begin{sketch}
At a high level, the proof proceeds in two steps. In {\bf Step 1}, we provide a sufficient condition (i.e., \eqref{eq:conditions}) for a profile to be problematic. Then in {\bf Step 2}, we construct a most equitable refinement for any profile that does not satisfy this condition. 

\myparagraph{Step 1.} To develop intuition behind the sufficient condition, let us revisit the proof of the ANR impossibility for $m=n=2$ in the Introduction. Indeed, the proof shows that $P= (1\succ 2, 2\succ 1)$ is a problematic profile under $\cor_\decspace$---the irresolute rule that always outputs $\{1,2\}$, because  any resolute rule $r$ (which  refines $\cor_\decspace$) fails anonymity or neutrality  at $P$ no matter how winners of other profiles are chosen. Let $\{d\} = r(P)$. The reason behind such failure is the existence of a permutation $\sigma$ over $\decspace$  such that
\begin{equation}
\label{eq:conditions}
\text{\bf (i) } d\ne \sigma(d) \text{,  and  \bf (ii) } \hist(P) = \sigma(\hist(P))
\end{equation}
In group theoretic terms, (i) says that $d$ is not a fixed-point of $\sigma$ and (ii) says that $\hist(P)$ is a fixed-point of $\sigma$. It is not hard to verify that for any profile and any rule $r$, if there exists a permutation that satisfies (i) and (ii), then ANR is guaranteed to fail at $P$ under $r$. 
In other words, the ANR impossibility holds if no decision is ``good'' at $P$. This notion of goodness is captured in the following definition, which defines ``good'' decisions. 
\begin{dfn}[{\bf\boldmath Fixed-point decisions}{}]
\label{dfn:fw} 
Given any $(\prefspace,\decspace)$ in the \commonsettings{}, for  any profile $P$, define {\em fixed-point decisions} at $P$ as: 
$$\fw(P)\triangleq  \{d\in \decspace: \forall \sigma\in\sgroup\text{ s.t. }\hist(\sigma(P)) = \hist(P), \sigma(d) = d\}$$
\end{dfn}

A similar observation holds for general irresolute rule $\cor$, with an additional constraint that $r$ must refine $\cor$. Following the same logic, it is not hard to verify that a profile $P$ is problematic under $\cor$  if for every $d\in \cor(P)$, there exists a permutation that satisfies \eqref{eq:conditions}. 
Formally, if $\fw(P)\cap\cor(P)=\emptyset$, then $P$ is a problematic profile under $\cor$. 

\myparagraph{Step 2.} We  explicitly construct  a refinement $r^*$ that satisfies $\anr$ at  all profiles $P$ such that $\fw(P)\cap\cor(P)\ne \emptyset$. Notice that a necessary condition for $r^*$ is, for any profile $P$ and  any profile  $P'$ that can be obtained from $P$ by applying a permutation over decisions and a permutation over the agents, $r^*(P')$ is determined by applying the same permutations to $r^*(P)$.  Therefore, for each {\em anonymous and neutral equivalence class (ANEC)}~\citep{Egecioglu2013:The-Impartial}---profiles that can be obtained from each other by applying permutations over alternatives and over agents---either all profiles satisfy $\anr$ or none of them satisfy $\anr$. 

Then, we show that the non-problematic profiles ($P$'s such that $\fw(P)\cap\cor(P)\ne \emptyset$) can be represented by unions of multiple ANECs. For each such ANEC, we first choose an arbitrary profile $P^*$ as the ``representative'' profile of the ANEC, then choose an arbitrary winner in $\fw(P^*)\cap\cor(P^*)$, and finally extend the winner to all profiles in the ANEC in a consistent way. The full version of the lemma and its full proof can be found in Appendix~\ref{app:proof-lem:anr}.
\end{sketch}
 
Following the proof of Lemma~\ref{lem:anr-characterization}, we immediately have the following characterization: a profile is problematic if and only if none of its co-winners is a fixed point decision.
\begin{prop}[{\bf Characterization of problematic profiles}{}]
\label{prop:problematic-profiles}
Given any $(\prefspace,\decspace)$ in the \commonsettings{} and any irresolute rule $\cor$ that satisfies anonymity and neutrality, for  any profile $P$,
\begin{center} $P$ is problematic $\Longleftrightarrow\fw(P)\cap\cor(P)= \emptyset$.
\end{center}
\end{prop}

We emphasize that the notion of ``most equitable'' in this paper is w.r.t. anonymity and neutrality, following their wide recognition as fundamental and uncontroversial notions of equity and fairness discussed in the Introduction.   Step 2 of the proof of Lemma~\ref{lem:anr-characterization} also characterizes all most equitable refinements, each of which uses different ``representative'' profiles and chooses different fixed point decisions.  

\section{The ANR Impossibilities}
\label{sec:chara}
To motivate the statement of the main theorem of this section (Theorem~\ref{thm:ANR-common}), which characterizes the ANR impossibility under the \commonsettings{}, let us revisit Moulin's condition~\citet{Moulin1983:The-Strategy} for the ANR impossibility under $(\prefspace,\decspace) = (\listset m,\listset 1)$, and reveal an equivalent condition, which we call  {\em partition condition}.
\begin{center}
\renewcommand{\arraystretch}{1.5}
\begin{tabular}{|c|c|c|}
\cline{1-1} \cline{3-3}
\bf Moulin's condition & & \bf Partition condition\\
\cline{1-1} \cline{3-3}
\begin{minipage}[c]{.2\textwidth}
$m$ is a sum of $n$'s non-trivial divisors.
\end{minipage}
& $\Longleftrightarrow$ &
\begin{minipage}[c]{.66\textwidth}
\vspace{1mm}
There exists a partition $\vec m$ of $m$ that satisfies\\
$\bullet$ {\bf sub-vector constraint:} no sub-vector of $\vec m$ sum up to $1$,\\
$\bullet$ {\bf change-making constraint:}   $n$ is feasible by  $\{\lcm(\vec m)\}$, where $\lcm(\vec m)$ is the least common multiplier of the elements in $\vec m$.
\vspace{1mm}
\end{minipage}\\
\cline{1-1} \cline{3-3}
\end{tabular}
\renewcommand{\arraystretch}{1}
\end{center}
The partition condition has two constraints. The first is on sub-vectors of  $\vec m$. The second requires that the following change-making problem has a solution: $n$ can be represented as a non-negative integer combination of a certain set of denominations with infinite supplies of coins.  The two conditions are equivalent because the sub-vector constraint is equivalent to no element of $\vec m$ being $1$, and  the change-making constraint  is equivalent to all elements of $\vec m$ being divisors of $n$.

%
%

Interestingly, Bubboloni and Gori~\citet{Bubboloni2014:Anonymous}'s condition for the ANR impossibility under $(\prefspace, \decspace)= (\listset{m},\listset m)$  also has an equivalent partition condition with a different  sub-vector constraint. 

\begin{center}
\renewcommand{\arraystretch}{1.5}
\begin{tabular}{|c|c|c|}
\cline{1-1} \cline{3-3}
\bf BG's condition & & \bf Partition condition\\
\cline{1-1} \cline{3-3}
\begin{minipage}[c]{.25\textwidth}
$m\ge$ $n$'s smallest non-trivial divisor.
\end{minipage}
& $\Longleftrightarrow$ &
\begin{minipage}[c]{.66\textwidth}
\vspace{1mm}
There exists a partition $\vec m$ of $m$ that satisfies\\
$\bullet$ {\bf sub-vector  constraint:} $\vec m$ contains less than $m$ $1$'s,\\
$\bullet$ {\bf change-making constraint:} $n$ is feasible by  $\{\lcm(\vec m)\}$.
\vspace{1mm}
\end{minipage}\\
\cline{1-1} \cline{3-3}
\end{tabular}
\renewcommand{\arraystretch}{1}
\end{center}
It turns out that, surprisingly, similar partition conditions characterize the  ANR impossibility under the \commonsettings{}. We now formally define the denominations that will be used in the change-making constraint. Given two vectors $\vec m$ and $\vec \ell$, we write $\vec m\ge \vec \ell$ or $\vec \ell\le \vec m$,  if $\vec m$ and $\vec \ell$ have the same length and  $\vec m$ is larger than or equal to  $\vec \ell$ element-wise.
\begin{dfn}[{$\lcmset{\vec m,\ell}{\circledast}$ and $\lcmset{\vec m,\ell}{\oslash}$}{}]
For any pair of  integers $\alpha\ge \beta\ge 0$, define 
\begin{equation*}
\label{eq:operations-dfn}
\alpha\circledast  \beta \triangleq \begin{cases} \alpha&\text{if }\beta>0\\1&\text{otherwise}\end{cases}  
\text{  and }  
\alpha\oslash \beta \triangleq \begin{cases} \frac{\lcm(\alpha,\beta)}{\beta}&\text{if }\beta>0\\1&\text{otherwise}\end{cases} 
\end{equation*}
Then, for any vector $\vec m\ge \vec1$ and $\ell\ge 1$,  define
\begin{align*}
&\lcmset{\vec m,\ell}{\circledast}\triangleq \left\{\lcm( \vec m\circledast \vec \ell):  \forall\  \vec 0\le \vec \ell\le \vec m \text{ and }\vec \ell\cdot \vec 1 = \ell\right\}\\
&\lcmset{\vec m,\ell}{\oslash}\triangleq  \left \{ \lcm( {\vec m}\oslash \vec\ell): \forall\  \vec 0\le \vec \ell\le \vec m \text{ and }\vec \ell\cdot \vec 1 = \ell\right\},
\end{align*}
where $\vec m\circledast \vec \ell$ and $\vec m\oslash \vec \ell$ are vectors obtained from element-wise applications of  $\circledast$ and $\oslash$, respectively.
\end{dfn}
For simplicity,  we assume that elements in $\vec m$ are sorted in non-increasing order. Elements in $\vec\ell$  are not sorted.
\begin{ex} 
\label{ex:coins} 
$(6,4)\le (8,6)$, $(8,6)\circledast (6,4) = (8,6)$, and $(8,6)\oslash (6,4)  = (\frac{\lcm(8,6)}{6},\frac{\lcm(6,4)}{4})= (4,3)$. 
When $m=4$, $\ell =2$, and $\vec m = (2,2)$, 
\begin{align*}
\lcmset{(2,2),2}{\circledast } = &\{\lcm((2,2)\circledast (2,0)), \lcm((2,2)\circledast (1,1)),\lcm((2,2)\circledast (0,2))\} = \{2\}\\
\lcmset{(2,2),2}{\oslash } = &\{\lcm((2,2)\oslash (2,0)), \lcm((2,2)\oslash (1,1)),\lcm((2,2)\oslash (0,2))\} = \{1,2\}
\end{align*}
Other partitions $\vec m$ of $m$, $\lcmset{\vec m,2}{\circledast}$, and $\lcmset{\vec m,2}{\oslash}$ are summarized below.
\begin{center}
\renewcommand{\arraystretch}{1.5}
\begin{tabular}{|r | *{5}{c|}}
\hline 
$\vec m=$ & $(4)$  & $(3,1)$ &  $(2,2)$ &  $(2,1,1)$  & $(1,1,1,1)$\\
\hline $\lcmset{\vec m, 2}{\circledast}=$ & $\{4\}$  & $\{3\}$ & $\{2\}$ & $\{1,2\}$ & $\{1\}$\\
\hline  $\lcmset{\vec m, 2}{\oslash}=$ & $\{2\}$  & $\{3\}$ & $\{1, 2\}$ & $\{1,2\}$ & $\{1\}$\\
\hline
\end{tabular}
\renewcommand{\arraystretch}{1}
\end{center} 


 

\end{ex}

\begin{thm}[{\bf\boldmath  ANR impossibility: \commonsettings{}}{}]
\label{thm:ANR-common}
For any $m\ge 2$, $n\ge 1$, $1\le \ell\le m$,  $1\le k\le m$, and any $(\prefspace,\decspace)$ in the \commonsettings{}, the ANR impossibility holds if and only if there exists a partition $\vec m$ of $m$ that satisfies
\begin{itemize}
\item [$\bullet$] {\bf  sub-vector  constraint:} 
$\begin{cases}
\text{$\vec m$ contains less than $k$ $1$'s}&\text{if }\decspace = \listset{k} \\
\text{no sub-vector of $\vec m$ sum up to $k$}&\text{if }\decspace = \committee{k}
\end{cases}$, and \\
\item [$\bullet$] {\bf  change-making constraint:} $n$ is feasible by  
$\begin{cases}
\lcmset{\vec m, \ell}{\circledast}&\text{if }\prefspace = \listset{\ell}\\
\lcmset{\vec m, \ell}{\oslash}&\text{if }\prefspace = \committee{\ell}
 \end{cases}$
 \end{itemize}
\end{thm}

%
Theorem~\ref{thm:ANR-common} naturally generalizes and strengthens previous characterizations. 
For example, when $\prefspace = \listset m $, we have $\ell=m$ and the only partition $\vec \ell $ of $\ell$ such that $\vec \ell\le \vec   m$ is $\vec \ell = \vec m$, which means that $\lcmset{\vec m, m}{\circledast} = \lcm(\vec m)$. Therefore,  applications of Theorem~\ref{thm:ANR-common} to $(\prefspace,\decspace) = (\listset m,\listset 1)$ and $(\prefspace,\decspace) = (\listset m,\listset m)$ give us the partition conditions that are equivalent to  Moulin~\citet{Moulin1983:The-Strategy}'s condition and Bubboloni and Gori~\citet{Bubboloni2014:Anonymous}'s condition  shown in the beginning of this section, respectively. See Example~\ref{ex:ANR-committee-m} in Appendix~\ref{sec:chara:notation-examples} for applications of Theorem~\ref{thm:ANR-common} to $\prefspace = \committee m$.

Let us look at another example where the preferences and decisions are both $\listset 2$.
\begin{ex}
\label{ex:list2-list2}
Let $m=4$ and $(\prefspace,\decspace) = (\listset 2, \listset 2)$. The set of partitions $\vec m$ of $m=4$ that satisfy the sub-vector constraint in Theorem~\ref{thm:ANR-common} is $\{(4), (3,1), (2,2)\}$. It follows from Theorem~\ref{thm:ANR-common} and the table in Example~\ref{ex:coins} that the ANR impossibility holds if and only if $n$ is feasible by $\{4\}$, $\{3\}$, or $\{2\}$, or equivalently, $2\mid n$ or $3\mid n$. For example, the ANR impossibility holds for $n=9$ but   not  for $n=7$. \\
\begin{minipage}[t][][b]{0.45 \textwidth} 
Similarly, we can characterize the ANR impossibility for other \commonsettings{} with $\ell=k=2$, which are summarized in the   table on the right.
\end{minipage} 
\hfill
\begin{minipage}[t][][b]{0.50\textwidth} 
\renewcommand{\arraystretch}{1}
\begin{tabular}{|c|*{2}{c|}}
 \hline
 \diagbox{$\prefspace$}{ANR Imp iff\ }{$\calD$} & $\listset 2$ & $\committee 2$ \\
 \hline
 $\listset 2$ & $2\mid n$ or $3\mid n$ &  $3\mid n$ or $4\mid n$  \\
\hline
 $\committee 2$ & every $n\in\mathbb N$ & $2\mid n$ or $3\mid n$ \\ 
\hline
 \end{tabular}
 \renewcommand{\arraystretch}{1} 
\end{minipage}

\end{ex}

\subsection{Proof Sketch of Theorem~\ref{thm:ANR-common}}

\begin{figure}[htp]
\centering
\includegraphics[width=\textwidth]{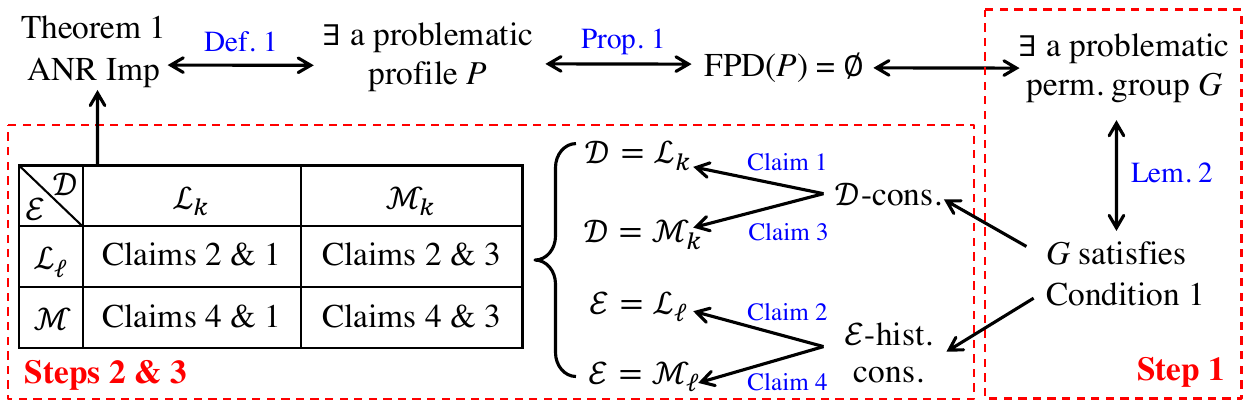}
\vspace{-4mm}
\caption{Overview of the proof of Theorem~\ref{thm:ANR-common}.\label{fig:proof-overview}}
\vspace{-4mm}
\end{figure}

The proof proceeds in three steps illustrated in Figure~\ref{fig:proof-overview}. 

In {\bf Step 1}, we define the notion of {\em problematic permutation group} (Definition~\ref{dfn:problematic-perm-group}), which is a sub-group of $\sgroup$ (all permutations over $\ma)$,  and proves that its existence is a necessary and sufficient for the ANR impossibility to hold (formally proved in Lemma~\ref{ANR-Cond} in Appendix~\ref{app:proof-thm:ANR-common}). In {\bf Step 2}, we focus on $(\prefspace,\decspace) = (\listset \ell, \listset k)$ and establish connections between the two constraints in Definition~\ref{dfn:problematic-perm-group}---the {\em $\decspace$ constraint} and the {\em $\prefspace$-histogram constraint}---and the sub-vector constraint and the change-making constraint  in the theorem statement, respectively (Claims~\ref{claim:D=Lk} and~\ref{claim:E=Ll}) . 
Finally, in {\bf Step 3}, we extend the connections to other preference and decision spaces in Claims~\ref{claim:D=Mk} and~\ref{claim:E=Ml}, and combine Claims~\ref{claim:D=Lk}--\ref{claim:E=Ml} to prove Theorem~\ref{thm:ANR-common} as shown in Figure~\ref{fig:proof-overview}.

\myparagraph{\bf Step 1: The ``problematic'' permutation group.} It follows from Proposition~\ref{prop:problematic-profiles} that the ANR impossibility holds if there exists a problematic profile $P$ under $\cor_\decspace$, i.e.,  $\fw(P)=\emptyset$. For any problematic profile $P$, consider the set of all permutations $\sigma$ to which $\hist(P)$ is invariant, i.e., $\sigma(\hist(P)) = \hist(P)$. That is,
$$G_P= \{\sigma\in\sgroup: \sigma(\hist(P)) = \hist(P)\}$$
It is not hard to verify that $G_P$ is a permutation group (i.e., a subgroup of $\sgroup$, the set of all permutations over $\ma$).  We call $G_P$ a {\em problematic permutation group}, because its existence (formally defined  without referring to a problematic profile) is equivalent to the existence of a problematic profile as we will see soon.  
For any set $X$,  let $\fixed_G(X)$ denote the set of all fixed points of $G$ in $X$, i.e., all $x\in X$ such that $g(x) =x$ holds for all $g\in G$. 
\begin{dfn}[{\bf Problematic permutation group}{}]
\label{dfn:problematic-perm-group}
A permutation group $G$ is {\em problematic} if it satisfies

$\bullet$ the {\bf $\decspace$ constraint:}  $\fixed_G(\decspace)=\emptyset $,

$\bullet$ the {\bf $\prefspace$-histogram constraint:}  $\fixed_G(\histset{m}{n}{\prefspace})\ne\emptyset $.
\end{dfn}
Notice that the first constraint requires that $G$ has no fixed point in $\decspace$, while the second constraint requires that $G$ has a fixed point in the histogram space.   To see why the existence of a problematic permutation group  implies the ANR impossibility, consider  any permutation group $G$ that satisfies the two conditions in Definition~\ref{dfn:problematic-perm-group}. Because of the $\prefspace$-histogram constraint, there exists a profile $P$ whose histogram is invariant to all permutations in $G$, and because of the $\decspace$ constraint, for every decision in $d\in\decspace$, there exists a permutation $\sigma$ in $G$ that maps $d$ to a different decision. It follows that $\fw(P)=\emptyset$, which means that $P$ is problematic, and therefore the ANR impossibility holds (Proposition~\ref{prop:problematic-profiles} applied to $\cor_\decspace$). The reverse direction also holds as proved in Lemma~\ref{ANR-Cond} in Appendix~\ref{app:proof-thm:ANR-common}.

\myparagraph{\bf Step 2: The $(\listset\ell,\listset k)$ case.} We establish connections between the two constraints in Definition~\ref{dfn:problematic-perm-group} and the sub-vector constraint and the change-making constraint in the theorem statement by considering the {\em orbits} of a permutation group $G$ on the set of alternatives $\ma$. An orbit in $\ma$  is a set of alternatives, each of which  can be obtained from each other by applying a permutation in $G$. 
\begin{ex}[{\bf An orbit in $\ma$}{}] 
\label{ex:orbit-main}
Let $m=5$ and $G=\{\id,(1,2), (3,4), (1,2)(3,4) \}$. Then there are three orbits: $\{1,2\}, \{3,4\}, \{5\}$. 
\end{ex}
It follows that the orbits constitute a partition of $\ma$, which means that the numbers of these orbits, called their {\em sizes} and are denoted by $\orbsize{G}{\ma}$ (formally defined in Definition~\ref{dfn:orbit-sizes}), is a vector that partitions $m$. For example, $\orbsize{G}{\ma} = (2,2,1)$ in Example~\ref{ex:orbit-main}. 

Then, we translate the two constraints in Definition~\ref{dfn:problematic-perm-group} by analyzing the orbit sizes of a problematic permutation, and then construct a problematic permutation group when the constraints in the theorem statement are satisfied. For example, when $(\prefspace, \decspace)= (\listset \ell,\listset k)$, the connections are formally established by the following two claims.

\appClaim{\bf\boldmath $\decspace = \listset k$}{claim:D=Lk}
{
For any permutation group $G$,  
$$\fixed_G(\listset k)=\emptyset \Longleftrightarrow \orbsize{G}{\ma} \text{ contains less than $k$ $1$'s}$$  
}
 
\appClaim{\bf\boldmath $\prefspace = \listset \ell$}
{claim:E=Ll}
{
For any permutation group $G$  and any partition $\vec m$ of $m$,
\begin{itemize}
\item [](i) $ \fixed_G(\histset{m}{n}{\listset \ell})\ne\emptyset \Rightarrow n$ is feasible by  $\lcmset{\orbsize{G}{\ma}, \ell}{\circledast}$ 
\item [](ii)  $n$ is feasible by  $\lcmset{\vec m, \ell}{\circledast}\Rightarrow \fixed_{G_{\vec m}}(\histset{m}{n}{\listset \ell})\ne\emptyset$.
\end{itemize}
}

In Claim~\ref{claim:E=Ll}, $G_{\vec m}$ is a special permutation group that is generated by the permutation that consists of cycles of alternatives whose sizes correspond to the elements in $\vec m$ (see Definition~\ref{dfn:perm-div} in Appendix~\ref{app:proof-thm:ANR-common}). For example, when $m=5$ and $\vec m =(2,2,1)$, we have $G_{\vec m} = \{\id, (1,2)(3,4)\}$. 

We are now ready to prove the $(\prefspace,\decspace) = (\listset \ell,\listset k)$ case of Theorem~\ref{thm:ANR-common} by combining Claim~\ref{claim:D=Lk} and Claim~\ref{claim:E=Ll}.  To prove the  {\bf \boldmath ``if'' direction}, suppose there exists a partition $\vec m$ of $m$ that satisfies the sub-vector constraint and the change-making constraint. Notice that $\orbsize{G_{\vec m}}{\ma} = \vec m$. Therefore, we have $\fixed_{G_{\vec m}}(\listset k)=\emptyset$ (the $\Leftarrow$ part of Claim~\ref{claim:D=Lk}) and $\fixed_{G_{\vec m}}(\histset{m}{n}{\listset \ell})\ne\emptyset$ (part (ii) of Claim~\ref{claim:E=Ll}). This means that there exists a permutation group (i.e., $G_{\vec m}$) that satisfies both constraints in Definition~\ref{dfn:problematic-perm-group}, which implies the ANR impossibility according to Lemma~\ref{ANR-Cond}. To prove the  {\bf \boldmath ``only if'' direction}, suppose the ANR impossibility holds. Then, by Lemma~\ref{ANR-Cond}, there exists a permutation group $G$ that satisfies  both constraints in  Definition~\ref{dfn:problematic-perm-group}. Then, $\orbsize{G}{\ma}$, which is a partition of $m$, satisfies the sub-vector constraint (the $\Rightarrow$ part of Claim~\ref{claim:D=Lk}) and the change-making constraint (part (i) of Claim~\ref{claim:E=Ll}).


A critical step in the proof of Claim~\ref{claim:E=Ll} is the application of the orbit-stabilizer theorem (see, for example~\cite[Theorem 2.65]{Sepanski2010:Algebra} and~\eqref{eq:orbit-stabilizer} in Appendix~\ref{app:proof-thm:ANR-common}), which reveals a connection between the number of alternatives in an orbit and the number of {\em stabilizers} of the orbit (permutations in $G$, to each of which all alternatives in the orbit are invariant). 

\myparagraph{\bf Step 3: Other \commonsettings{}.} We extend the proof in Step 2 to other \commonsettings{} by proving Claim~\ref{claim:D=Mk} (for $\decspace = \committee k$) and Claim~\ref{claim:E=Ml} (for $\prefspace = \committee \ell$), and combining Claims~\ref{claim:D=Lk}-\ref{claim:E=Ml} in ways similar to Step 2, as shown in Figure~\ref{fig:proof-overview}. The proofs involve multiple novel applications of the orbit-stabilizer theorem. The full proof can be found in Appendix~\ref{app:proof-thm:ANR-common}.

\subsection{At-Large ANR Impossibilities}
In many applications,  $(\prefspace,\decspace)$ is determined before  $n$ is known, and $n$ is often large. Therefore,  it is important to understand whether the ANR impossibility holds for every sufficiently large $n$. To this end, we introduce a notion called {\em at-large ANR impossibility}.
\begin{dfn}[\bf At-large ANR impossibility]
Given   $(\prefspace,\decspace)$, we say that 
 {\em at-large ANR impossibility} holds, if there exists $N$ such that the ANR impossibility holds for every $n>N$.
\end{dfn}
In other words, if at-large ANR impossibility does not hold, then there exist infinitely many $n$'s such that ANR rules exist. 

\begin{thm}[{\bf At-large ANR impossibility: up-to-$L$ preferences}{}] 
\label{thm:at-large-up-to-L}
For any $(\prefspace,\decspace)\in \{\listset{\le L}, \committee{\le L}:1\le L\le m\}\times \{\listset{k}, \committee{k}:1\le k\le m\}$, at-large ANR impossibility holds if and only if 
\begin{center}
\renewcommand{\arraystretch}{1}
\begin{tabular}{|c|*{2}{l|}}
 \hline
 \diagbox{$\prefspace$} {$\calD$} & \hspace{23mm}$\listset k$ & \hspace{23mm}$\committee k$ \\
 \hline
 $\listset {\le L}$ &
 \begin{tabular}{@{}l@{}} (i) $m=5$, or \\
 (ii) $m\ge 7$, or \\(iii) $k \ge 2$  \end{tabular} 
 &  
 \begin{tabular}{@{}l@{}} (i) $m=5$ and $k\le m-1$, or \\
 (ii) $m\ge 7$ and $k\le m-1$, or \\(iii) $2\le k\le m-2$  \end{tabular}\\
\hline
$\committee {\le L}$
&
\begin{tabular}{@{}l@{}}  
 (i) $m=5$, or\\
 (ii) $m\ge 7$, or \\
 (iii) $\max(L,k)\ge 2$,  \\
 except $(m,L,k)\in\{ (2,1,2),  (3,2,1)\}$ 
 \end{tabular} & 
\begin{tabular}{@{}l@{}}
 (i) $m=5$ and $k\le m-1$, or\\
  (ii) $m\ge 7$ and $k\le m-1$, or\\
 (iii) $[L]\not\subseteq \{k,m-k\}$ and $k\le m-1$\\
 \ \\
\end{tabular} \\ 
\hline
 \end{tabular}
 \renewcommand{\arraystretch}{1}
\end{center}
Moreover, under the conditions in the table, the ANR impossibility holds for every $n\ge m^2/2$.
\end{thm}
\begin{sketch}  
We prove the  $(\listset {\le L},\listset k)$ case to illustrate the idea.   The proof of other cases can be found in Appendix~\ref{sec:proof-at-large-up-to-L}. We first prove in Claim~\ref{claim:suffice-ell=1} in Appendix~\ref{sec:proof-at-large-up-to-L} that at-large ANR impossibility holds for any $L$ if and only if it holds for $L=1$, meaning that it suffices to focus on the $L=1$ case. 

To prove the ``if'' direction, for each combination of $m$ and $k$  in the theorem statement, we specify a partition $\vec m$ of $m$  that satisfies the sub-vector constraint, and specify one or two  $\vec \ell$'s so that either a coin has value $1$ or two coins  are co-primes.  When $k\ge 2$, let $\vec m = (m-1,1)$ and $\ell = (0,1)$. Then, $1\in \lcmset{\vec m,1}{\circledast}$.  When $m=5$ and $k=1$, let $\vec m = (3,2)$ and $\vec \ell\in \{(0,1),(1,0)\}$. Then, $\{2,3\} \subseteq \lcmset{\vec m,1}{\circledast}$.  When $m\ge 7$ and $k=1$, let $\vec m$ be defined as in \eqref{eq:vecd} and let $\ell\in \{(0,1),(1,0)\}$. 
\begin{equation} 
\vec m = (m_1,m_2)\triangleq \begin{cases}
\left(\frac{m+1}{2},\frac{m-1}{2}\right)& 2\nmid m\\
\left(\frac{m}{2}+1,\frac{m}{2}-1\right)& 4\mid m\\
\left(\frac{m}{2}+2,\frac{m}{2}-2\right)&4\nmid  m \text{ and }  2\mid m
\end{cases}\hspace{10mm}
\label{eq:vecd}
\end{equation}
Then $\lcmset{\vec m,1}{\circledast}$ contains two co-prime numbers. 

The ``only if'' direction covers the cases where $k=1$ and $m\in \{2,3,4,6\}$, and  is proved by enumerating all $\vec m$'s that satisfy the sub-vector constraint, which tell us all $n$'s for which the ANR impossibility does not hold, as summarized below.  
\begin{center}
\renewcommand{\arraystretch}{1.3}
\begin{tabular}{|r | *{2}{c|}}
\hline 
$m$ & partitions satisfying the sub-vector constraint  &  ANR Imp does not hold for\\
\hline 
$6$ & $\{(6), (4,2), (3,3),(2,2,2)\}$ & $2\nmid n$ and $3\nmid n$\\
\hline 
$4$ & $\{(4), (2,2)\}$ & $2\nmid n$\\
\hline $3$ & $\{(3) \}$ & $3\nmid n$\\
\hline $2$ & $\{(2) \}$ & $2\nmid n$\\
\hline
\end{tabular}
\renewcommand{\arraystretch}{1}
\end{center}  
\end{sketch}

\begin{ex}[{\bf NYC Democratic mayoral primary election}{}]
\label{ex:NYC}
In this election, there were $13$ qualified candidates. Each voter can rank up to five candidates.  This corresponds to the setting where $m=13$, $n$ is close to a million, and $(\prefspace,\decspace) = (\listset{\le 5}, \listset 1)$. The ANR impossibility holds under this setting, because according to Theorem~\ref{thm:at-large-up-to-L}, at-large ANR impossibility holds. Notice that $n\ge m^2/2$. Therefore, the ANR impossibility holds as well.
\end{ex}

\section{The Most-Favorable-Permutation Tie-Breaking}
\label{sec:alg}
Lemma~\ref{lem:anr-characterization} only guarantees the existence of a most equitable refinement of $\cor$. 
In this section, we propose a novel tie-breaking mechanism to obtain a most equitable refinement, and then design a polynomial-time algorithm to compute it when  $\prefspace = \listset m$.

The idea behind our tie-breaking  mechanism is the following. As shown in  the proof of Lemma~\ref{lem:anr-characterization}, the key step in obtaining a most equitable refinement is to identify and fix a ``representative'' profile for each equivalent class. However, the number of equivalent classes is exponentially large in $m$, which means that  pre-computing the representative profiles  takes exponential time. This is the main challenge identified in previous work for $(\prefspace,\decspace)=(\listset m,\listset 1)$~\citep{Bubboloni2021:Breaking}. To address this challenge, we define a priority order over the histograms by lexicographically extending a priority order $\rhd$  over $\ma$, and then choose a representative profile of an equivalent class to be one whose histogram has the highest priority in $\rhd$. 


\begin{dfn} [{\bf Lexicographic extensions of $\rhd$}{}]
\label{dfn:ext-priority}
Let  $\rhd\in \listset m$. For any $i\le m$, We extend $\rhd$ to $ \listset i$ (respectively, $  \committee i$), such that $i$-lists (respectively, $i$-committees) are compared lexicographically w.r.t.~their top-ranked alternatives, second-ranked alternatives (respectively, most-preferred alternative, second-most-preferred alternatives),  etc.  We extend $\rhd$ to
 $\histset{m}{n}{\prefspace}$, such the vectors in $\histset{m}{n}{\prefspace}$ are  compared lexicographically, favoring histograms with higher values in more important coordinates according to $\rhd$.   
\end{dfn}
W.l.o.g.~we let $\rhd = [1\succ\cdots\succ m]$ in this paper. In all examples, the coordinates in $\histset{m}{n}{\prefspace}$ are ordered in the increasing order according to $\rhd$. 

\begin{ex}
\label{ex:rhd} 
Continuing the setting of Example~\ref{ex:perm}, among the $8$ types of rankings in  $\sigma_1( P^{+} )$ and $ \sigma_2(P^{+}) $, we have 
 $12345 \rhd 12435 \rhd  21345 \rhd 21435 \rhd  31425\rhd 32415\rhd 41325 \rhd 42315$.  
Therefore, $\sigma_1(P^{+})\rhd \sigma_2(P^{+})$. 
\end{ex} 
Next, for any profile $P$, we define a class of permutations that map $\hist(P)$ to a histogram with the highest priority according to $\rhd$.
\begin{dfn}[{\bf Most favorable permutations (MFPs)}{}] For any profile $P$, 
\begin{equation}
\label{eq:mfp}
\hp(P)\triangleq \arg\max\nolimits^{\rhd}_{\sigma\in\sgroup}\hist(\sigma(P))
\end{equation}
\end{dfn}
Note that $\arg\max$ maximizes the priority of $\hist(\sigma(P))$ according $\rhd$ instead of maximizing the rank number of $\hist(\sigma(P))$  in $\rhd$ (which corresponding to minimizing its priority). We are now ready to introduce the tie-breaking mechanism. The high-level idea is, for any profile $P$, we first find a ``representative'' histogram by applying a most favorable permutation $\sigma_\hp$, then use $\rhd$ to break ties among fixed-point decisions, and finally map the sole winner back via $\sigma_\hp^{-1}$. This is equivalent to choosing a fixed-point decision that has the highest priority in $\rhd$ after applying $\sigma_\hp$.  

\begin{dfn}[{\bf  MFP tie-breaking}{}]
\label{dfn:mfpb} Given a ``backup'' tie-breaking mechanism $\tb$, a profile $P$,  and $D\subseteq \calD$, let $\sigma_\hp \in \hp(P)$  be an arbitrary most favorable permutation. Define
\begin{equation}
\label{eq:mfp-tb}
\hpp_\tb(P,D) \triangleq \begin{cases}
\arg\max\nolimits^{\rhd}_{d\in \fw(P)\cap D} \sigma_\hp(d)& \text{if }\fw(P)\cap D\ne\emptyset\\
\tb(P,D) &\text{otherwise}
\end{cases}
\end{equation}
\end{dfn}
The following theorem confirms that $\hpp_\tb$ is well-defined, in the sense that the choice of  $\sigma_\hp$ does not matter, and $\hpp_\tb$ outputs a most equitable refinement. 

\begin{thm}
\label{thm:MFP} Under the \commonsettings{}, for any anonymous and neutral rule $\cor$ and any backup tie-breaking mechanism $\tb$, $\hpp_\tb$ is well-defined and $\breaking{\hpp_\tb}{\cor}$ is a most equitable refinement.
\end{thm}
\begin{sketch}
The high-level idea naturally follows the idea in the proof of Lemma~\ref{lem:anr-characterization}. For each profile $P$, we identify the representative profile in its ANEC by applying $\sigma_\hp$. Then, $r(P)$ is determined by permuting the winner under $\sigma_\hp(P)$ via $\sigma_\hp^{-1}$. The full proof can be found in Appendix~\ref{sec:proof-thm:MFP}.
\end{sketch}

\begin{ex}[{\bf\boldmath  $\hpp_\tb$}{}]
\label{ex:mfp}
Continuing the setting of Examples~\ref{ex:perm} and  let $\sigma_1^* =(1,4,2,3)$. Then, $\hp(P^{+}) = \{\sigma_1, \sigma_1^*\}$ and $\sigma_1(P^{+}) =\sigma_1^*(P^{+})$.  Let $\cor=\veto$.  For any backup tie-breaking mechanism $f$, we have $\hpp_\tb( P^{+},\veto(P^{+}))  = \hpp_\tb( P^{+},\{1,2,3,4\}) = \{3\}$, because 
$$\sigma_1(3) = \sigma_1^*(3) = 1 \rhd 2 =\sigma_1(4) = \sigma_1^*(4)$$
\end{ex}

Next, we  present a polynomial-time algorithm (Algorithm~\ref{alg:tie-breaking}) to compute $\hpp_\tb$ under the  \commonsettings{} where $\prefspace = \listset m$. The main idea is that  any MFP   must map $R\in\listset m$ with the highest multiplicity in $\hist(P)$ to the linear order with the highest priority w.r.t.~$\rhd$, which is $\rhd$ itself (as a linear order in $\listset m$). 
Then, a MFP can be computed by exploring all such permutations and choosing the one that maps $P$ to the profile whose histogram has the highest priority in $\rhd$.

Recall that for any profile $P$ and any ranking $R$,  $[\hist(P)]_R$ is the $R$-component of the histogram of $P$.    Let $\text{MPR}(P)\triangleq \arg\max_{R} [\hist(P)]_R$  denote the set of all {\em most popular rankings} in $P$, i.e., the rankings with the highest multiplicity in $P$.  Algorithm~\ref{alg:tie-breaking} stores $\hist(P)$ as lists of linear orders  that appear at least once in $P$ and their multiplicities.

\begin{algorithm}[htp]
\caption{{\bf (MFP tie-breaking)} Compute $\hpp_\tb(P,D)$ for $P=(R_1,\ldots,R_n)\in (\listset m)^n$ and $D\subseteq \calD$.}\label{alg:tie-breaking}
\begin{algorithmic}[1] 
\STATE  Let $\sigma_{jj'}$ be the permutation such that $\sigma_{jj'}(R_j) = R_{j'}$. Compute  
\label{step:perm}
\begin{minipage}{\textwidth-6mm}
\begin{equation}
\label{equ:unc-comp}
\stab(\hist(P)) = \bigcap\nolimits_{j\le n}\{\sigma_{jj'}:1\le j'\le n\text{ such that }[\hist(P)]_{R_j} = [\hist(P)]_{R_{j'}}\} 
\end{equation}
\end{minipage}
\STATE Compute $\fw(P)$, and  {\bf if} $\fw(P)\cap D = \emptyset$,  {\bf then return} $\tb(P,D)$. \label{step:fixed}
\STATE {\bf  for} every most popular ranking $R\in \text{MPR}(P)$, compute   $\sigma_R\in\sgroup$ such that $\sigma_R(R) = \ \rhd $.
\label{step:mpr}
\STATE Compute $R^* \in \arg\max_{R\in \text{MPR}(P)}^\rhd  \sigma_R(\hist(P))$.\label{step:sigmastar}
\STATE  \Return $\arg\max_{d\in \fw(P)\cap D}^\rhd  \sigma_{R^*}(d)$.
\label{step:return}
\end{algorithmic}
\end{algorithm}
$\stab(\hist(P))$ in step~\ref{step:perm} of Algorithm~\ref{alg:tie-breaking} is the set of {\em stabilizers} of $\hist(P)$ in $\sgroup$, i.e., all permutations $\sigma$ such that $\sigma(\hist(P))=\hist(P)$.
\begin{ex}[{\bf Execution of Algorithm~\ref{alg:tie-breaking}}{}]
\label{ex:alg}
We run Algorithm~\ref{alg:tie-breaking} on $P^{+}$ and $\cor=\veto$ in the setting of Examples~\ref{ex:perm}  and~\ref{ex:mfp}.  {\bf In step~\ref{step:perm}}, the right-hand side of Equation~\eqref{equ:unc-comp} for $j$ such that $R_j=43125$ is $\{\id , (1,2)\}$, because $[\hist(P^+)]_{43125}=[\hist(P^+)]_{ 43215} =1$. According to Equation~\eqref{equ:unc-comp}, 
$\stab(\hist(P^{+})) = \{\id , (1,2)\}$. 
{\bf In step~\ref{step:fixed}}, we have $\fw(P^{+})\cap \veto(P^{+})= \{3,4\}$.  
{\bf In step~\ref{step:mpr}}, $\text{MPR}(P)=\{34125, 34215, 43125,43215\}$. Let $\sigma_2^* \triangleq (1,4)(2,3)$. We have $\sigma_{34125} = \sigma_1$ (defined in Example~\ref{ex:perm}), $\sigma_{34215} = \sigma_1^*$ (defined in Example~\ref{ex:mfp}), $\sigma_{43125} = \sigma_2$ (defined in Example~\ref{ex:perm}), and $\sigma_{43215} = \sigma_2^*$. 
Then {\bf in step~\ref{step:sigmastar}}, we have $\sigma_1(P^{+}) = \sigma_1^*(P^{+})$ and  $\sigma_2(P^{+}) = \sigma_2^*(P^{+})$. Recall from Example~\ref{ex:rhd} that $\sigma_1(P^{+}) \rhd \sigma_2(P^{+})$. Therefore, we can choose $R^* = 34125$. Finally, in {\bf step~\ref{step:return}},
$\sigma_{R^*}(3) = 1 \rhd 2 = \sigma_{R^*}(4)$, which means that $(\hpp_\tb\ast\veto)(P^{+})=\{3\}$.
\end{ex}
The correctness and  efficiency of Algorithm~\ref{alg:tie-breaking} are guaranteed by the following theorem, whose proof can be found in Appendix~\ref{app:alg}. 
\begin{thm}
\label{thm:hpp-opt}
For any polynomially computable $f$ and any $(\listset m,\decspace)$ in the \commonsettings{}, Algorithm~\ref{alg:tie-breaking}  computes $\hpp_\tb$ in polynomial time. 
\end{thm}  
For any $(\prefspace,\decspace)$ in the \commonsettings{} where $\prefspace\ne \listset m$, computing a MFP of $P$ according to Equation~\eqref{eq:mfp} takes $O(m!\cdot \text{poly}(m,n))$ time by enumerating all permutations $\sigma$, which means that computing $\hpp_\tb$ takes $O(m!\cdot \text{poly}(m,n))$ time if $f$ is polynomial-time computable. This idea works for even more general settings and is formally presented as Algorithm~\ref{alg:MFP-general} in Appendix~\ref{sec:general-MFP}.

\section{Related work and discussions}
{\bf Circumventing the ANR impossibility.} As commented by Zwicker~\citet{Zwicker2016:Introduction}, using tie-breaking mechanisms is a common approach towards circumventing the ANR impossibility. Most previous work focuses on $\prefspace=\listset m$, while other settings have gain popularity recently. For example,  the characterization of ANR impossibility is an open question for approval-based committee voting, where $(\prefspace,\decspace) = (\committee \ell, \committee k)$~\cite[Chapter 3.1]{Lackner2023:Multi-Winner}. Our results contribute to the fundamental understandings of the tie-breaking approach: we now know how to break ties  optimally (Lemma~\ref{lem:anr-characterization} and Theorem~\ref{thm:MFP}) and computationally efficiently (Theorem~\ref{thm:hpp-opt}). 

\myparagraph{Characterizations of ANR rules.}  A series of previous work focused on (fully or partially) characterizing the existence of ANR rules by conditions on $m$  and $n$, mostly under   certain \commonsettings{} where $\prefspace = \listset m$. 

When {\boldmath $(\prefspace,\decspace) = (\listset m,\listset 1)$},  Moulin~\citet{Moulin1983:The-Strategy} proved that an ANR rule   exists if and only if $m$ cannot be represented as a sum of $n$'s non-trivial (i.e., $>1$)  divisors. Moulin~\citet{Moulin1983:The-Strategy} also proved that  an ANR rule that also satisfies Pareto efficiency exists  if and only if $m$ is  smaller than  $n$'s smallest non-trivial divisor, which is equivalent to $\gcd(m!,n)=1$, where $\gcd$ represents the greatest common divisor.  Campbell and Kelly~\citet{Campbell2015:The-finer} pointed out that sometimes ANR comes at the cost of other desirable properties. Dogan and Giritligil~\citet{Dogan2015:Anonymous}  characterized ANR rules that also satisfy monotonicity. Ozkes and Sanver~\citet{Ozkes2021:Anonymous} investigated the existence of rules that satisfy anonymity, resolvability, and a weaker notion of neutrality. 

When {\boldmath $(\prefspace,\decspace) = (\listset m,\listset m)$}, Bubboloni and Gori~\citet{Bubboloni2014:Anonymous} proved that an ANR rule exists if and only if   $\gcd(m!,n)=1$. Bubboloni and Gori~\citet{Bubboloni2015:Symmetric}  considered a setting where voters are divided into groups and investigated a weaker notion of in-group anonymity.  

When {\boldmath $(\prefspace,\decspace) = (\listset m,\committee k)$}, Bubboloni and Gori~\citet{Bubboloni2016:Resolute} proved that an ANR rule that satisfies Pareto efficiency exists if and only if $\gcd(m!,n)=1$. Bubboloni and Gori~\citet{Bubboloni2021:Breaking}  provided sufficient conditions for the existence of ANR rules that satisfy weaker notions of anonymity and neutrality. However, the conditions for the ANR impossibility to hold under  $(\prefspace,\decspace) = (\listset m,\listset k)$ as well as under other \commonsettings{} not mentioned above remain open, which are resolved by our Theorem~\ref{thm:ANR-common}. 

Other related works include the seminar work by May~\citep{May52:Set}, which investigated anonymous and neutral rules for two alternatives where ties are allowed, and characterized the majority rule. Bartholdi et al.~\citet{Bartholdi2021:Equitable} proposed a new and weaker notion of equity in the same voting setting as May's setting~\citep{May52:Set}  that justifies representative democracy. 

\myparagraph{Tie-breaking mechanisms.} ANR is satisfied for any irresolute rule $\cor$ at  any profile $P$ under which there are no ties. Therefore, some previous work focused on designing tie-break mechanisms to preserve as much ANR satisfaction as possible. When {\boldmath $(\prefspace,\decspace) = (\listset m,\listset 1)$}, Dogan and Giritligil~\citet{Dogan2015:Anonymous} proposed to use an ANR ranking rule to break ties  when $\gcd(m!,n)=1$. Bubboloni and Gori~\citet{Bubboloni2016:Resolute} characterized conditions for the existence of a tie-breaking mechanism to preserve weaker versions of anonymity and neutrality.  Xia~\citet{Xia2020:The-Smoothed} showed that the agenda tie-breaking and the fixed-voter tie-breaking are far from being optimal under a large class of semi-random models, proposed a tie-breaking mechanism, and proved that it is asymptotically optimal for even $n$'s. Our MFP tie-breakings are optimal as they computes most equitable refinements, which achieve highest ANR satisfaction regardless of the underlying distribution over profiles. 

When {\boldmath $(\prefspace,\decspace) = (\listset m,\listset m)$ or $(\listset m,\committee k)$}, Bubboloni and Gori~\citet{Bubboloni2021:Breaking} proposed   tie-breaking mechanisms for $k$-committees and rankings that are based on defining a system of ``representatives'' for each equivalent class of profiles. They also highlighted the computational challenges in their approach: (using our notation) ``{\em finding explicitly a system of representatives is in general a complex problem. However, that can be managed for small values of $m$ and $n$, that is, when the size $(m!)^n$ of $(\listset{m})^n$ is not too large}''.  It remained an open question of whether ANR can be optimally preserved by some polynomial-time tie-breaking mechanism when $\prefspace = \listset m$, which is address by our MFP  breakings (Algorithm~\ref{alg:tie-breaking} and Theorem~\ref{thm:hpp-opt}). 

\section{Future Work}
There are many directions for future work. For example, immediate open questions include characterizing conditions for the ANR impossibility for other commonly studied settings, such as partial orders; designing efficient algorithms for computing MFP; characterizing likelihood of ANR violations; analyzing randomized tie-breaking mechanisms; studying the likelihood of ANR violations under natural statistical models; verifying the likelihood of ANR violations on real-world data. Our general theory presented in Appendix~\ref{sec:general} provides some useful tools for addressing these questions. More generally, how to extend the study to  other desirable properties together with ANR (or other notions of fairness), such as Pareto efficiency and monotonicity, is another important direction for future work.


\bibliographystyle{splncs04}
\bibliography{/Users/administrator/GGSDDU/references}
 \newpage
 \setcounter{tocdepth}{2}
\tableofcontents
\newpage
\appendix
\section{Additional Preliminaries}
\label{app:prelim}

\subsection{Commonly Studied Voting Rules}

\paragraph{\bf Weighted Majority Graphs.}  For any profile $P$ and any pair of alternatives $a,b$, let $ P[a\succ b]$ denote the total weight of votes in $P$ where $a$ is preferred to $b$. Let $\wmg(P)$ denote the {\em weighted majority graph} of $P$, whose vertices are $\ma$ and whose weight on edge $a\ra b$ is $w_P(a,b) = P[a\succ b] - P[b\succ a]$. 

A voting rule is said to be {\em weighted-majority-graph-based (WMG-based)} if its winners only depend on the WMG of the input profile. We consider the following commonly studied WMG-based rules. For each rule, we will define its $\calD =\listset k$ version for $1\le k\le m-1$, and its $\committee k$ version chooses all (unordered) alternatives involved in its $\listset k$ version. Recall that $\calD = \ma$ is a special case of $\calD =\listset k$, where $k=1$; and $\calD = \ml(\ma)$ is a special case of $\calD =\listset k$, where $k=m$. 
\begin{itemize}
\item {\bf Copeland.} The Copeland rule is parameterized by a number $0\le \alpha\le 1$, and is therefore denoted by Copeland$_\alpha$, or $\copeland$ for short. For any fractional profile $P$, an alternative $a$ gets $1$ point for each other alternative it beats in their head-to-head competition, and gets $\alpha$ points for each tie. Copeland$_\alpha$  for $\calD = \listset k$  chooses all linear extensions of weak orders of top $k$ alternatives w.r.t.~their Copeland scores. In other words, for any profile $P$, $a_1\succ \cdots\succ a_k\in\copeland(P)$ if and only if (1) for any $k_1,k_2 \le k$ such that $a_{k_1}\succ a_{k_2}$, the Copeland score of $a_{k_2}$ is  no more than the Copeland score of $k_1$; and (2) for any $k'>k$, the Copeland score of $a_{k'}$ is  no more than the Copeland score of $k_2$.
\item  {\bf Maximin.} For each alternative $a$, its min-score  is defined to be $\min_{b\in\ma}w_P(a,b)$. Maximin for $\calD=\listset k$ chooses all linear extensions of weak orders of top $k$ alternatives w.r.t.~their maximin scores.
\item {\bf Ranked pairs.} Given a profile $P$, we fix  edges in $\wmg(P)$ one by one in a non-increasing order w.r.t.~their weights (and sometimes break ties), unless it creates a cycle with previously fixed edges. After all edges (with positive, $0$, or negative weights) are considered, the fixed edges represent  a linear order over $\ma$ and the ranked top-$k$ alternatives are chosen. We consider all tie-breaking methods to define the ranked top-$k$ alternatives. This is known as the {\em parallel-universes tie-breaking (PUT)}~\citep{Conitzer09:Preference}.
\item {\bf Schulze.} For any directed path in the WMG, its strength is defined to be the minimum weight on any single edge along the path. For any pair of alternatives $a,b$, let $s[a,b]$ be the highest weight among all paths from $a$ to $b$. Then, we write $a\succeq b$ if and only if $s[a,b]\ge s[b,a]$, and 
the strict version of this binary relation, denoted by $\succ$, is transitive~\citep{Schulze11:New}. The Schulze rule for $\calD=\listset k$ chooses ranked top-$k$ alternatives in all linear extensions of  $\succeq$. 
\end{itemize}

\noindent{\bf STV.} The single transferable vote (STV) rule has $m-1$ rounds. In each round, the  alternative with the lowest plurality score is eliminated. STV for $\calD=\listset k$ chooses ranked top-$k$ alternatives according to the inverse elimination order. Like ranked pairs, PUT is used to select the winning $k$-lists. 


\subsection{Group Theory}
\label{sec:group-theory}

Let $|G|$ denote the number of elements in $G$, also called the {\em order} of $G$. A group $H$ is a {\em subgroup} of another group $G$, denoted by $H\leqslant G$, if $H\subseteq G$ and the operation for $H$ is the restriction of the operation for $G$ on $H$.

For any $g\in G$ in a group and any $K\in \mathbb N$, we let $g^K \triangleq \underbrace{g\circ\cdots\circ g}_K$. The {\em symmetric group} over $\ma=[m]$, denoted by $\sgroup$, is the set of all permutations over $\ma$. A permutation $\sigma$ that maps each $a\in \ma$ to $\sigma(a)$ can be represented in two ways.
\begin{itemize}
\item {\em Two-line form:} $\sigma$ is represented by a $2\times m$ matrix, where the first row is $(1,2,\ldots, m)$ and the second row is $(\sigma(1),\sigma(2),\ldots, \sigma(m))$. 
\item {\em Cycle form:} $\sigma$ is represented by non-overlapping cycles over $\ma$, where each cycle $(a_1,\cdots, a_K)$ represent $a_{i+1}=\sigma(a_i)$ for all $i\le K-1$, and with $a_{1}=\sigma(a_K)$. 
\end{itemize}
For example, all permutations in $S_3$ are represented in two-line form and cycle form respective in the Table~\ref{tab:s3}.
\begin{table}[htp]
\centering
\begin{tabular}{|@{\small\ }c|@{} c@{}|@{}c@{}|@{}c@{}|@{}c@{}|@{}c@{}|@{}c@{}|}
\hline
Two-line&$\left(\begin{array}{ccc} 1& 2 & 3 \\ 1 & 2 & 3
\end{array}\right)$ & $\left(\begin{array}{ccc} 1& 2 & 3 \\ 2 &1 & 3
\end{array}\right)$& $\left(\begin{array}{ccc} 1& 2 & 3 \\ 1 & 3 & 2
\end{array}\right)$& $\left(\begin{array}{ccc} 1& 2 & 3 \\ 3 & 2 & 1
\end{array}\right)$& $\left(\begin{array}{ccc} 1& 2 & 3 \\ 2 & 3 & 1
\end{array}\right)$& $\left(\begin{array}{ccc} 1& 2 & 3 \\ 3 & 1 & 2
\end{array}\right)$ \\
\hline
Cycle& $()$ or $\id$& $(1,2)$& $(2,3)$& $(1,3)$& $(1,2,3)$& $(1,3,2)$ \\
\hline
\end{tabular}
\caption{${\cal S}_{\ma}$ with $m=3$.\label{tab:s3}}

\end{table}

A {\em permutation group} $G$ over $\ma$ is a subgroup of $\sgroup$, i.e., $G\leqslant \sgroup$.  For any vector $\vec d$ of non-negative integers that sum up to $m$,  we define a special permutation $\sigma_{\vec d}$ and a permutation group $G_{\vec d}$ generated by $\sigma_{\vec d}$ as follows.


\section{Materials for Section~\ref{sec:MER}}

\subsection{Group-Theoretic Definition of Fixed-Point Decisions}
\begin{dfn}[{\bf\boldmath Fixed-point decisions}{}]
\label{dfn:fw} 
Given any $(\prefspace,\decspace)$ in the \commonsettings{}, for  any profile $P\in \prefspace^n$ and any set of decisions $D\subseteq\decspace$, define
\begin{align*}
\text{\bf\boldmath Stabilizers of $\hist(P)$: } &\stab(\hist(P)) \triangleq \{\sigma\in\sgroup: \hist(\sigma(P)) = \hist(P)\}, \\ 
\text{\bf Fixed-point decisions: }&\end{align*}
$$\fw(P)\triangleq  \fixed_{\stab(\hist(P))}(\calD) \triangleq \{\sigma\in\sgroup: \hist(\sigma(P)) = \hist(P)\}$$
\end{dfn}
In words, a stabilizer of $\hist(P)$ is a permutation $\sigma$ under which $\hist(P)$ is invariant. A fixed point of $\stab(\hist(P))$ in $\decspace$ is a decision that is invariant under all stabilizers of $\hist(P)$. And a fixed-point decision is simply a decision in $D$ that is also a fixed point of $\stab(\hist(P))$. For example, in the setting of Example~\ref{ex:problematic-veto}, $\stab(\hist(P^-)) = \{\id, \sigma_{(1,2)}\}$, where $\id$ is the identity permutation that does not change any alternative, $\fixed_{\stab(\hist(P^-))} = \{3,4,5\}$,  $\fw(P^-,\veto(P^-)) =  \{3,4,5\}\cap \{1,2\} = \emptyset$. 

\subsection{Lemma~\ref{lem:anr-characterization} and Its Proof}
\label{app:proof-lem:anr}

\appLem{\bf Existence of most equitable refinements}{lem:anr-characterization}{
Under \commonsettings{}, any anonymous and neutral irresolute rule  has a most equitable refinement.  Moreover, for every most equitable refinement $r^*$ and every $P\notin \problematic{\cor}{n}{\prefspace}{\decspace}$, $ r^*(P) \subseteq \fw(P)\cap\cor(P)$.
}
 \begin{proof} 
The   lemma is proved in the following two steps.

\myparagraph{Step 1: any profile $P$ such that $\fw(P)\cap \cor(P) =\emptyset $ is in $\problematic{\cor}{n}{\prefspace}{\decspace}$.} The proof is similar to the reasoning in Example~\ref{ex:problematic-veto}. Suppose for the sake of contradiction that  there exists $P$ with $\fw(P)\cap \cor(P) =\emptyset $ and a refinement $r$ of $\cor$ such that $\anr(r,P)=1$. Let $r(P) = \{d\}$. Then, $d$ is not a fixed point of $\stab(\hist(P))$, which means that there exists a permutation $\sigma\in \stab(\hist(P))$ such that $\sigma(d)\ne d$. Because $\hist(\sigma(P)) =\sigma(\hist(P)) = \hist(P)$, by anonymity, we have $r(\sigma(P)) = r(P) = \{d\}$, but by neutrality, we have $r(\sigma(P)) = \sigma(r(P)) = \{\sigma(d)\}\ne \{d\}$, which is a contradiction.

\myparagraph{Step 2: there exists a refinement $r^*$ such that for all  profiles $P$ such that $\fw(P)\cap \cor(P) \ne \emptyset $, $\anr(r^*,P) = 1$.} The proof is constructive and is similar to the idea in the following example.
\begin{ex}[{\bf\boldmath Non-problematic profile}{}]
\label{ex:good-veto} Let $P^{+}$ be the profile  in Example~\ref{ex:perm}. We have $ \veto(P^{+}) = \{1,2,3,4\}$ and $\stab(\hist(P^+)) = \{\id, \sigma_{(1,2)}\}$, which means that $\fixed_{\stab(\hist(P^{+}))}(\decspace) = \{3,4,5\}$.  Hence $\fw(P^+)\cap\veto(P^+) = \{3,4\}$. Let $r$ be a refinement such that $r(P^+) = \{3\}$, and for every
permutation $\sigma$ and every profiles $P$ such that $\hist(P)=\hist(\sigma(P^+))$, let $r(P) = \{\sigma(3)\}$. Then, we have $\anr(r,P^+) = 1$. A similar construction works if we choose $r(P^+) = \{4\}$, but if $r(P^+)\subseteq \{1,2\}$, then $\anr(r,P^+) = 0$ by considering $\sigma_{(1,2)}$ as in Example~\ref{ex:problematic-veto}.
\end{ex} 
Formally, Step 2 consists of the following three steps. 

{\bf Step 2.1.~Define an equivalence relationship.}  Let $\sim$ denote the equivalence relationship in $\histset{m}{n}{\prefspace}$, such that for any pair of profiles $P_1,P_2$, $P_1\sim P_2$ if and only if there exists a permutation $\sigma$ such that $\hist(P_1) =\hist(\sigma(P_2))$.

{\bf Step 2.2.~Choose ``representative'' profiles.} For each equivalent class (of $n$-profiles) according to $\sim$, we arbitrarily choose a ``representative'' profile $P$, fix it throughout the proof, and define $r^*(P)$ to be  an arbitrary but fixed decision  in $\fw(P)\cap \cor(P)$.

{\bf Step 2.3.~Extend to other profiles.} For any ``representative'' profile $P$ and any $P'\sim P$, let $\sigma$ denote the permutation such that $\hist(P') = \sigma(\hist(P))$. We define $r^*(P) = \sigma(r^*(P))$.

 It follows that for every $P$ such that $\fw(P)\cap \cor(P) \ne \emptyset $, $\anr(r^*,P) = 1$.  Moreover, if there exits a most equitable refinement $r^*$ and $P\notin \problematic{\cor}{n}{\prefspace}{\decspace}$ such that $ r^*(P) = \{d\}\nsubseteq \fw(P)\cap\cor(P)$, then $d$ is not a fixed point under $\stab(\hist(P))$, which means that there exists a permutation $\sigma\in \stab(\hist(P))$ such that $\sigma(d)\ne d$. Notice that $\sigma(\hist(P))=\hist(P)$. Therefore, anonymity or neutrality is violated at $P$ under $r^*$, which is a contradiction.
\end{proof}
\section{Materials for Section~\ref{sec:chara}}

\subsection{Notation and Examples}
\label{sec:chara:notation-examples}
\begin{dfn}[{\bf Feasible numbers}{}]
\label{dfn:feasible}
Given a set $\calC = \{n_1,\ldots,n_T\}$ of positive integers, we let $\intcomb{\calC}$ denote the set of all positive integers that can be represented as non-negative linear combinations of elements in $\calC$. That is, let $\vec n = (n_1,\ldots, n_T)$, define
$$\intcomb{\calC}\triangleq \left\{\vec\alpha\cdot\vec n:\vec\alpha\in {\mathbb Z}_{\ge 0}^T\text{ and }\vec\alpha\ne \vec 0\right\}$$
\end{dfn}
In other words, $n\in \intcomb{\calC}$ if and only if $n$ is feasible by $\calC$.

Next, we define $\vdl{m}{k}$ and  $\vda{m}{k}$ to be the two sets of partitions of $m$ that satisfy the sub-vector constraint for $\decspace = \listset k$ and $\decspace = \committee k$ in the statement of Theorem~\ref{thm:ANR-common}, respectively. Let $\vpl{m,n}{\ell} $ and  $\vpa{m,n}{\ell} $ denote the partitions of $m$ that satisfy the change-making constraint for $n$ for $\prefspace = \listset \ell$ and $\prefspace = \committee \ell$ in the statement of Theorem~\ref{thm:ANR-common}, respectively. More precisely, we have the following definition.

\begin{dfn}
\label{dfn:partitions} 
Given any $m,n,\ell$, and $k$, we define
\begin{align*}
\vdl{m}{k}\triangleq & \{\vec m: \text{$\vec m$ contains less than $k$ $1$'s}\}\\
 \vda{m}{k}  \triangleq & \{\vec m: \text{no sub-vector of $\vec m$ sum up to $k$}\}\\
\vpl{m,n}{\ell}   \triangleq & \{\vec m: \text{$n$ is feasible by $\lcmset{\vec m, \ell}{\circledast}$}\} \\
\vpa{m,n}{\ell} \triangleq & \{\vec m: \text{$n$ is feasible by $\lcmset{\vec m, \ell}{\oslash}$}\}
\end{align*}
\end{dfn} 

We assume that all vectors are represented in non-increasing order of their components. For example, $\vdl{4}{2} = \{(4), (3,1), (2,2)\}$. 

\begin{ex}
\label{ex:ANR-committee-m}
When $\prefspace = \committee m$, i.e., $\ell = m$, for any partition $\vec m$ of $m$, we have $\lcmset{\vec m, \ell}{\oslash} = \lcmset{\vec m, m}{\oslash} = \{\lcm(\vec m\oslash \vec m)\} = \{1\}$. Therefore, the ANR impossibility holds (for any $n$) if and only if there exists an $\vec m$ that satisfies the sub-vector constraint.

That is, according to Theorem~\ref{thm:ANR-common},  for all $(\committee m,\decspace)$ in the \commonsettings{}, except $\decspace = \committee m$, the ANR impossibility holds for all $n\ge 1$. It is not hard to verify that this is true because all voters can only cast the same vote ($\ma$), and when $\decspace \ne \committee m$, for any decision, there exists a permutation that maps it to a different decisions, which proves that the ANR impossibility holds.

When $(\prefspace,\decspace) = (\committee m, \committee m)$, according to Theorem~\ref{thm:ANR-common}, the ANR impossibility theorem does not hold. It is easy to verify that this is true because there is only one decision $\ma$, and any permutation maps it to itself. 
\end{ex}

\subsection{Proof of Theorem~\ref{thm:ANR-common}}
\label{app:proof-thm:ANR-common}

\appThm{\bf\boldmath   ANR impossibility: \commonsettings{}}{thm:ANR-common}
{ 
For any $m\ge 2$, $n\ge 1$, $1\le \ell\le m$,  $1\le k\le m$, and any $(\prefspace,\decspace)$ in the \commonsettings{}, the ANR impossibility holds if and only if there exists a partition $\vec m$ of $m$ that satisfies
\begin{itemize}
\item [$\bullet$] {\bf  sub-vector  constraint:} 
$\begin{cases}
\text{$\vec m$ contains less than $k$ $1$'s}&\text{if }\decspace = \listset{k} \\
\text{no sub-vector of $\vec m$ sum up to $k$}&\text{if }\decspace = \committee{k}
\end{cases}$, and \\
\item [$\bullet$] {\bf  change-making constraint:} $n$ is feasible by  
$\begin{cases}
\lcmset{\vec m, \ell}{\circledast}&\text{if }\prefspace = \listset{\ell}\\
\lcmset{\vec m, \ell}{\oslash}&\text{if }\prefspace = \committee{\ell}
 \end{cases}$
 \end{itemize}
 }

\begin{proof}
{\bf Overview.} In Step 1, we introduce the notion of problematic permutation groups, which are permutation groups that satisfy two group theoretic constraints (Definition~\ref{dfn:problematic-perm-group}) and prove that the existence of a problematic permutation group characterizes the ANR impossibility. Then in Step 2, we show that the two constraints are equivalent to the sub-vector constraint and the change-making constraint under \commonsettings{}, respectively. We present the full proof for  $(\prefspace,\decspace) = (\listset\ell,\listset k)$, where the equivalences are shown in Claim~\ref{claim:D=Lk} (for $\decspace = \listset k$) and Claim~\ref{claim:E=Ll} (for $\prefspace = \listset \ell$), respectively. Finally,  other \commonsettings{} are handled in Step 3.

\myparagraph{Step 1: Group theoretic constraints.} As discussed in the beginning of Section~\ref{sec:chara}, the ANR impossibility holds if and only if there exists a problematic $n$-profile $P$ under $\cor_\decspace$, and by Lemma~\ref{lem:anr-characterization}, 
\begin{equation}
\label{eq:fixed-empty}
\fw( P, \cor_\decspace(P))=\decspace\cap \fixed_{\stab(\hist(P))}(\calD) = \fixed_{\stab(\hist(P))}(\calD)=\emptyset
\end{equation}
Notice that $\stab(\hist(P))$ is a   subgroup of $\sgroup$ with two properties. First, as \eqref{eq:fixed-empty} shows, $\stab(\hist(P))$ does not have a fixed point in $\decspace$. Second,  $\stab(\hist(P))$ has at least one fixed point, e.g.,  $\hist(P)$, in  $\histset{m}{n}{\prefspace}$. This motivates us to define the following condition and prove that it characterizes the ANR impossibility. We recall that $\fixed_G(X)$ consists of fixed points of $G$ in $X$, i.e., all $x\in X$ such that $g(x) =x$ holds for all $g\in G$. (Also see Definition~\ref{dfn:stab} in Appendix~\ref{sec:general-MFP} for its formal definition.)

Recall from  Definition~\ref{dfn:problematic-perm-group}  that a problematic permutation group is characterized by the following two constraints:

$\bullet$ the {\bf $\decspace$ constraint:}  $\fixed_G(\decspace)=\emptyset $,

$\bullet$ the {\bf $\prefspace$-histogram constraint:}  $\fixed_G(\histset{m}{n}{\prefspace})\ne\emptyset $.

\begin{lem}
\label{ANR-Cond}
The ANR impossibility holds if and only if there exists a problematic permutation group. 
\end{lem}
\begin{proof} We have already proved the ``only if'' part  by letting $G = \stab(\hist(P))$, where $P$ is a problematic profile. To prove the ``if'' direction, let $G$ denote the permutation group that satisfies both constraints in  Definition~\ref{dfn:problematic-perm-group} and let $P$ denote any profile such that $\hist(P)$ satisfies the $\prefspace$-histogram constraint for $G$. It follows that $G\subseteq \stab(\hist(P))$, which means that $\fixed_{\stab(\hist(P))}(\decspace)\subseteq \fixed_G(\decspace) = \emptyset$ (due to the $\decspace$ constraint for $G$). 
\end{proof}

The remainder of the proof  establishes  the equivalence between the two constraints in  Definition~\ref{dfn:problematic-perm-group} and the sub-vector constraint and the change-making constraint in the theorem statement under the \commonsettings{}. This is achieved by a group theoretic approach that analyzes the sizes of orbits in $\ma$ under $G$, formally defined as follows, where $G\leqslant \sgroup$ means  that $G$ is a subgroup of $\sgroup$. 
\begin{dfn}[{\bf\boldmath Orbit sizes}{}]
\label{dfn:orbit-sizes}
For any $G\leqslant \sgroup$ and any set $X$ that permutations over $\ma$ can be naturally applied (formal defined as ``acts  on'' in Definition~\ref{dfn:group-action} in Appendix~\ref{sec:dfn-general}), the {\em orbit} of $x\in X$ under $G$ is defined to be $\orbit_G(x)\triangleq \{g(x):g\in G\}$. Let $\orbsize{G}{X}$ denote the vector that represents the sizes of non-overlapping orbits in $\ma$ under $G$ in  non-increasing order.
\end{dfn}
Because $\ma$ can be partitioned to orbits under $G$, $\orbsize{G}{\ma}$ is a partition of $m$.
\begin{ex}[{\bf\boldmath Orbit sizes}{}] Let $m=4$, $n=2$, $(\prefspace, \decspace) = (\listset {1},\listset 2)$, and $G = \{\id, (1,2) (3,4)\}$. We have $\orbsize{G}{\calD} = \orbsize{G}{\listset {2}} = (2,2,2,2,2,2)$, because $\listset {2}$ can be partitioned into the following $6$ orbits under $G$:
\begin{align*}\listset {2} = \{1\succ 2, 2\succ 1\}&\cup \{1\succ 3, 2\succ 4\}\cup \{1\succ 4, 2\succ 3\}\cup \{3\succ 1, 4\succ 2\}\\
&\cup \{3\succ 2, 4\succ 1\}\cup \{3\succ 4, 4\succ 3\}
\end{align*}
$\orbsize{G}{\histset{m}{n}{\prefspace}} =    (2,2,2,2,1,1)$, because $\histset{m}{n}{\prefspace}$ can be partitioned into  $6$ orbits under $G$, i.e., $O_1,\ldots,O_6$ in the table below. Each column after the first represents a histogram, where $0$ entries are omitted:
\begin{center}
\begin{tabular}{|c|c:c|c:c|c:c|c:c|c|c|}
\hline  $\decspace = \listset 1 $ & \multicolumn{10}{c|}{$\histset{m}{n}{\prefspace}$}\\
\hline
$1$ &  $2$ &  & & & $1$ & & $1$ & & $1$& \\
$2$ &   & $2$ & & &   &$1$ &  & $1$ & $1$& \\
$3$ &   &  & $2$& &   & $1$& $1$ &  & & $1$\\
$4$ &   &  & & $2$& $1$  & &  & $1$ & & $1$\\
\hline \multicolumn{1}{c|}{}&\multicolumn{2}{c|}{$O_1$}&\multicolumn{2}{c|}{$O_2$}&\multicolumn{2}{c|}{$O_3$}&\multicolumn{2}{c|}{$O_4$}& $O_5$& $O_6$\\
\cline{2-11}
\end{tabular}
\end{center}
For example, $O_1 = \{(2,0,0,0), (0,2,0,0)\}$. $\orbsize{G}{ \listset 1} = (2,2)$, because $\ma= \listset 1$ is partitioned into   two orbits: $\listset 1 = \{1,2\} \cup \{3,4\}$.
\end{ex}

\myparagraph{Step 2: the $(\prefspace,\decspace) = (\listset \ell,\listset k)$ case.}  The proof proceeds in three steps. First, for $\decspace = \listset k$, we reveal a relationship (Claim~\ref{claim:D=Lk}) between the $\decspace$ constraint in  Definition~\ref{dfn:problematic-perm-group} and the sub-vector constraint in the statement of the theorem. Second,  for $\prefspace = \listset \ell$, we reveal a relationship (Claim~\ref{claim:E=Ll}) between the $\prefspace$-histogram constraint in  Definition~\ref{dfn:problematic-perm-group}  and the change-making constraint in the statement of the theorem. Finally, we combine the two claims to prove the theorem.
  
\begin{claim}[{\bf\boldmath $\decspace = \listset k$}{}]
\label{claim:D=Lk}
For any $G\leqslant \sgroup$ and any $1\le k\le m$,  
$$\fixed_G(\listset k)=\emptyset \Longleftrightarrow \orbsize{G}{\ma} \text{ contains less than $k$ $1$'s}$$  
\end{claim}
\begin{proof} 
We first prove the ``$\Leftarrow$'' direction. Suppose $\orbsize{G}{\ma} $ contains less than $k$ $1$'s and suppose for the sake of contradiction that $G$ has a fixed point in $R\in \listset k$, and suppose $R$ is a linear order over $A\subseteq \ma$ with $|A| = k$. Then, for every $a\in A$ and every $\sigma\in G$, we have  $\sigma (a) = a$, or in other words, every $a\in A$ is a fixed point of $G$ on $\ma$, which means that $\orbsize{G}{\ma} $ contains at least $k$ $1$'s (corresponding to the alternatives in $A$), which  is a contradiction.

The  ``$\Rightarrow$'' direction proved by proving its contraposition: if $\orbsize{G}{\ma}$ contains at least $k$ $1$'s, then $\fixed_G(\listset k)\ne \emptyset$.   Let $A\subseteq \ma$ denote any set of $k$ alternatives whose corresponding components in $\orbsize{G}{\ma}$ are $1$'s, or equivalently, for every $a\in A$, we have $|\orbit_{G}(a)|  = 1$. Let $R$ denote an arbitrary ranking over $A$. It follows that $R\in \listset k$ and for all $\sigma\in G$, $\sigma(R) = R$, which means that $R$ is a fixed point under $G$ and therefore, $\fixed_G(\listset k)\ne \emptyset$.
\end{proof}

For $\prefspace = \listset\ell$, the relationship between the $\prefspace$-histogram constraint in  Definition~\ref{dfn:problematic-perm-group} and the change-making constraint is weaker than Claim~\ref{claim:D=Lk}, in the sense that the $\Leftarrow$ direction only holds for a special type of permutation groups $G_{\vec m}$  defined as follows.
\begin{dfn}
\label{dfn:perm-div}
Given any partition $\vec m = (m_1,\ldots, m_T)$ of $m$, define
$$
\sigma_{\vec m} \triangleq(\underbrace{1,\ldots,m_1}_{m_1}) (\underbrace{m_1+1,\ldots, m_1+m_2}_{m_2}) \cdots(\underbrace{m-m_T+1,\ldots,m}_{m_T})$$ 
$$G_{\vec m}\triangleq \{(\sigma_{\vec m})^{K}: K \in \mathbb N\}
$$
\end{dfn}
That is, $\sigma_{\vec m}$  consists of cyclic permutations among groups of alternatives whose sizes are $m_1,\ldots, m_T$, respectively.  $G_{\vec m}$ is the {\em cyclic group} generated by $\sigma_{\vec m}$.  We have $|G_{\vec m}|=\lcm(\vec m)$.

\begin{claim}[{\bf\boldmath $\prefspace = \listset \ell$}{}]
\label{claim:E=Ll}
For any $G\leqslant \sgroup$, any $1\le \ell\le m$,  and any partition $\vec m$ of $m$,
\begin{itemize}
\item [(i)] $ \fixed_G(\histset{m}{n}{\listset \ell})\ne\emptyset \Rightarrow n$ is feasible by  $\lcmset{\orbsize{G}{\ma}, \ell}{\circledast}$ 
\item [(ii)]  $n$ is feasible by  $\lcmset{\vec m, \ell}{\circledast}\Rightarrow \fixed_{G_{\vec m}}(\histset{m}{n}{\listset \ell})\ne\emptyset$.
\end{itemize}
\end{claim}
\begin{proof} {\bf Part (i).} 
Choose any $\vec h\in  \fixed_G(\histset{m}{n}{\listset \ell})$.   Let $O_1,\ldots, O_T$ denote the orbits in $\listset \ell$ under $G$ such that each $\ell$-list in each orbit appears at least once in $\vec h$. That is,
$$\forall 1\le t\le T,\forall R\in O_t, [\vec h]_R >0$$
Because $\vec h$ is a fixed point of   $G$, the $\ell$-lists in the same obit $O_t$ appear the same number of times in $\vec h$. For every $1\le t\le T$, fix an arbitrary $\ell$-list $R^*_t\in O_t$. We have $\sum_{t=1}^T|O_t|\times [\vec h]_{R^*_t} = n$. Therefore, $n\text{ is feasible by } {\{|O_1|,\ldots, |O_T|\}}$.

To prove that $n$ is feasible by  $\lcmset{\orbsize{G}{\ma}, \ell}{\circledast}$, it suffices to prove that each number in $\{|O_1|,\ldots, |O_T|\}$ is coarser than some denomination $\lcm(\orbsize{G}{\ma}\circledast \vec \ell)$ in $\lcmset{\orbsize{G}{\ma}, \ell}{\circledast}$. That is, we will prove that for every $  t\le T$, there exist a sub-vector $\vec m_t$ of $\orbsize{G}{\ma}$ that satisfies
\begin{itemize}
\item [] {\bf constraint (a):} $\vec m_t\cdot\vec 1 \ge \ell$, 
\item [] {\bf constraint (b):} $\lcm(\vec m_t)$ is a divisor of $|O_t|$. 
\end{itemize}
If such $\vec m_t$ exists, then we let $\vec\ell_t$ denote an arbitrary partition of $\ell$ whose elements are positive only if they correspond to elements in $\vec m_t$. It follows that $\lcm(\orbsize{G}{\ma}\circledast \vec \ell_t)$ is a divisor of $\lcm(\vec m_t)$, which is a divisor of $|O_t|$. 

We explicitly construct $\vec m_t$ as follows. Let $R_t\in O_t$ denote an arbitrary $\ell$-list. For every orbit $O$ in $\ma$ under $G$, if any alternative in $O$ appears in $R_t$, then $\vec m_t$ has a component $|O|$. More precisely, let $O_1^t, O_2^t, \ldots, O_S^t$ denote the orbits in $\ma$ under $G$ that touches $R_t$. W.l.o.g., suppose  $|O_1^t|\ge \cdots\ge  |O_S^t|$. Then,  $\vec m_t = (|O_1^t|,\ldots, |O_S^t|)$. 

 By construction, $\vec m_t\cdot\vec 1  $ is at least the number of alternatives that appear in $R_t$, which is $\ell$.  This means that $\vec m_t$ satisfies constraint (a). To see that $\vec m_t$ satisfies constraint (b), it suffices to prove that for every $1\le s\le S$, $|O_s^t|$ is a divisor of $|O_t|$.  Let $a_s$ denote an alternative  that appears in both $R_t$ and $|O_s^t|$.  We will prove  constraint (b) by applying the {\em orbit-stabilizer theorem} (see, e.g.,~\cite[Theorem 2.65]{Sepanski2010:Algebra}) twice, one to the orbit $O_s^t$ of $a_t$ under $G$ and the other to the orbit $O_t$ of $R_t$ under $G$. 

\begin{tcolorbox}
{\bf Orbit-stabilizer theorem. }{\em For any $x$ in the set that $G$ acts on, 
 \begin{equation}
 \label{eq:orbit-stabilizer}
 |G| = |\stab_G(x)|\times |\orbit_G(x)|
 \end{equation}
 }
 \end{tcolorbox}
 
Notice that $O_s^t = \orbit_G(a_t) = \{g(a_t): g\in G\}$ and $\stab_G(a_t) = \{g\in G: g(a_t)=a_t\}$ is the set of stabilizers of $a_t$. Therefore, by the orbit-stabilizer theorem, we have $|G| = |\stab_G(a_t)|\times |O_s^t|$. Also  notice   that $O_t = \orbit_G(R_t)$ is the orbit of $R_t$ under $G$ and $\stab_G(R_t)\le G$ is the set of stabilizers of $R_t$. Therefore, by the orbit-stabilizer theorem, we have  $|G|=|\stab_G(R_t)|\times |O_t|$. Then,
\begin{equation}
\label{eq:GaR}|G| = |\stab_G(a_t)|\times |O_s^t| = |\stab_G(R_t)|\times |O_t|
\end{equation}
Notice that any stabilizer $\sigma\in \stab_G(R_t)$  maps every alternative that appear in $R_t$, particularly $a_t$, to itself. Therefore, $\sigma \in \stab_G(a_t)$, which means that $\stab_G(R_t)$ is a subgroup of $\stab_G(a_t)$. By Lagrange's theorem, $|\stab_G(R_t)|$ is a divisor of $\stab_G(a_t)$. This observation, combined with \eqref{eq:GaR}, implies that $|O_s^t| $ is a divisor of $|O_t|$. Therefore, $\vec m_t$ satisfies constraint (b). 

{\bf Part (ii).} We will explicitly construct a profile $P\in({\listset \ell})^n$ whose histogram is a fixed point of $G_{\vec m}$. Let $\vec m = (m_1,\ldots,m_T)$  be the partition of $m$ such that $n$ is feasible by $\lcmset{\vec m,\ell}{\circledast}$. Let $\ma = A_1\cup A_2\cdots\cup A_T$ denote a partition of $\ma$ such that for every $t\le T$,  $|A_t| = m_t$. Moreover, let $\vec\ell_1, \ldots, \vec\ell_S$  be non-negative partitions of $\ell$ such that $n$ is feasible by $\{\lcm(\vec m\circledast \vec\ell_s):s\le S\}$.
 More precisely, let $\alpha_1,\ldots,\alpha_S$ be non-negative integers such that
$n = \sum\nolimits_{s=1}^S \alpha_s\cdot \lcm(\vec m\circledast \vec\ell_s)$.

Next, we define the $\ell$-lists that will be used to construct the profile $P$. For every $s\le S$, let $R_s$ denote an arbitrary $\ell$-list over an arbitrary subset of $\ell$ alternatives of  
\begin{equation}\label{eq:unionA}
\bigcup\nolimits_{t\le T: [\vec\ell_s]_t\ne 0} A_t
\end{equation}
That is,  $R_s$ involves alternatives  in the union of  $A_t$'s for all $t$ such that the $t$-th element of $\vec\ell_s$ is strictly positive. $R_s$ is well-defined, because $\vec\ell_s\cdot\vec 1 =\ell$ and $\vec\ell_s \le \vec m$.

Consider the orbits of $R_1,\ldots,R_S$ under $G_{\vec m}$. It follows that for every $s\le S$, $|\orbit_{G_{\vec m}}(R_s)|$  divides $\lcm(\vec m\circledast \vec\ell_s)$, because for every alternative $a$ in \eqref{eq:unionA}, suppose $a\in A_t$, then we have $(\sigma_{\vec m})^{|A_t|}(a) = a$. Notice that $|A_t|$ divides $\lcm(\vec m\circledast \vec\ell_s)$. Therefore,  $(\sigma_{\vec m})^{\lcm(\vec m\circledast \vec\ell_s)}(a) = a$, which means that  $(\sigma_{\vec m})^{\lcm(\vec m\circledast \vec\ell_s)}(R_s) = R_s$. 

Finally, we define the following profile
$$P  \triangleq \bigcup\nolimits_{s=1}^S\left(\alpha_s\cdot\frac{\lcm(\vec m\circledast \vec\ell_s)}{|\orbit_{G_{\vec m}}(R_s)|}\right)\times \orbit_{G_{\vec m}}(R_s)$$

\begin{ex}
Let $m=5$, $n=7$, $\ell =2$, $\vec m=(3,2)$, $\vec\ell_1 = (2,0)$ and $\vec\ell_2=(0,2)$. Then, $n=7=\lcm(\vec m\circledast \vec \ell_1)+2\times \lcm(\vec m\circledast \vec \ell_1)$.  $\sigma_{\vec m} = (1,2,3) (4,5)$, and we can let $\ma =\underbrace{\{1,2,3\}}_{A_1}\cup \underbrace{\{4,5\}}_{A_2}$, $R_1 = [1\succ 2]$ and $R_2 = [4\succ 5]$. 
Then, $ \orbit_{G_{\vec m}}(R_1) = \{1\succ 2, 2\succ 3, 3\succ 1\}\text{ and }\orbit_{G_{\vec m}}(R_2) = \{4\succ 5,5\succ 4\}$.   $P = \{1\succ 2, 2\succ 3, 3\succ 1\}\cup 2\times \{4\succ 5,5\succ 4\}$.
\end{ex}

Then, for every $s\le S$ and every $\sigma\in G_{\vec m}$, $\hist(\sigma(\orbit_{G_{\vec m}}(R_s))) = \hist(\orbit_{G_{\vec m}}(R_s))$. Therefore, $\hist(P)$ is a fixed point of $G_{\vec m}$, which completes the proof.  
\end{proof}

We are now ready to prove the $(\prefspace,\decspace) = (\listset \ell,\listset k)$ case of Theorem~\ref{thm:ANR-common} by combining Claim~\ref{claim:D=Lk} and Claim~\ref{claim:E=Ll}.  To prove the  {\bf \boldmath ``if'' direction}, suppose there exists a partition $\vec m$ of $m$ that satisfies the sub-vector constraint and the change-making constraint. Notice that $\orbsize{G_{\vec m}}{\ma} = \vec m$. Therefore, we have $\fixed_{G_{\vec m}}(\listset k)=\emptyset$ (the $\Leftarrow$ part of Claim~\ref{claim:D=Lk}) and $\fixed_{G_{\vec m}}(\histset{m}{n}{\listset \ell})\ne\emptyset$ (part (ii) of Claim~\ref{claim:E=Ll}). This means that there exists a permutation group (i.e., $G_{\vec m}$) that satisfies both constraints in Definition~\ref{dfn:problematic-perm-group}, which implies the ANR impossibility according to Lemma~\ref{ANR-Cond}. To prove the  {\bf \boldmath ``only if'' direction}, suppose the ANR impossibility holds. Then, by Lemma~\ref{ANR-Cond}, there exists a permutation group $G$ that satisfies  both constraints in  Definition~\ref{dfn:problematic-perm-group}. Then, $\orbsize{G}{\ma}$, which is a partition of $m$, satisfies the sub-vector constraint (the $\Rightarrow$ part of Claim~\ref{claim:D=Lk}) and the change-making constraint (part (i) of Claim~\ref{claim:E=Ll}).

\myparagraph{Step 3: Other $(\prefspace,\decspace)$ settings.} For $\prefspace = \committee \ell$ and $\decspace  = \committee k$, we prove the following counterparts to Claim~\ref{claim:D=Lk} and Claim~\ref{claim:E=Ll}, respectively.

\begin{claim}[{\bf\boldmath $\decspace = \committee k$}{}]
\label{claim:D=Mk} 
For any $G\leqslant \sgroup$ and any $1\le k\le m$,  
$$\fixed_G(\committee k)=\emptyset \Longleftrightarrow \orbsize{G}{\ma} \text{ has no sub-vector that sum up to $k$}$$ 
\end{claim}

\begin{claim}[{\bf\boldmath $\prefspace = \committee \ell$}{}]
\label{claim:E=Ml}
For any $G\leqslant \sgroup$, any $1\le \ell\le m$,  and any partition $\vec m$ of $m$,
\begin{itemize}
\item [(i)] $ \fixed_G(\histset{m}{n}{\committee \ell})\ne\emptyset \Rightarrow n$ is feasible by  $\lcmset{\orbsize{G}{\ma}, \ell}{\oslash}$ 
\item [(ii)]  $n$ is feasible by  $\lcmset{\vec m, \ell}{\oslash}\Rightarrow \fixed_{G_{\vec m}}(\histset{m}{n}{\committee \ell})\ne\emptyset$.
\end{itemize}
\end{claim}
The proof of Claim~\ref{claim:D=Mk} is similar to the proof of Claim~\ref{claim:D=Lk}
while the proof of Claim~\ref{claim:E=Ml} is more complicated and involves multiple novel applications of the orbits-stabilizer theorem. The full proofs can be found in Appendix~\ref{sec:proof-claim:D=Mk} and Appendix~\ref{sec:proof-claim:E=Ml}, respectively.   Then,    other $(\prefspace,\decspace)$ settings in Theorem~\ref{thm:ANR-common}  are proved in a similar way as the $(\listset\ell,\listset k)$ case, using combinations of Claims~\ref{claim:D=Lk}, \ref{claim:E=Ll}, \ref{claim:D=Mk}, and~\ref{claim:E=Ml}  shown in the following table.

\begin{center}
\renewcommand{\arraystretch}{1}
\begin{tabular}{|c|*{2}{c|}}
 \hline
 \diagbox{$\prefspace$}{Proved by }{$\calD$} & $\listset k$ & $\committee k$ \\
 \hline
 \rule{0pt}{15pt}$\listset \ell$ & Claims~\ref{claim:E=Ll}\&\ref{claim:D=Lk} & Claims~\ref{claim:E=Ll}\&\ref{claim:D=Mk}  \\[5pt]
\hline
\rule{0pt}{15pt}$\committee \ell$ &  Claims~\ref{claim:E=Ml}\&\ref{claim:D=Lk} & Claims~\ref{claim:E=Ml}\&\ref{claim:D=Mk} \\[5pt]
\hline
 \end{tabular}
 \renewcommand{\arraystretch}{1}
\end{center} 
\end{proof}

\subsection{Proof of Claim~\ref{claim:D=Mk}}
\label{sec:proof-claim:D=Mk}

\appClaim{{\bf\boldmath $\decspace = \committee k$}}{claim:D=Mk}
{For any $G\leqslant \sgroup$ and any $1\le k\le m$,  
$$\fixed_G(\committee k)=\emptyset \Longleftrightarrow \orbsize{G}{\ma} \text{ has no sub-vector that sum up to $k$}$$ }
 
\begin{proof} We first prove the ``$\Leftarrow$'' direction. Suppose $\orbsize{G}{\ma} $ does not contain a sub-vector whose elements sum up to $k$. Suppose for the sake of contradiction that $G$ has a fixed point in  $\committee k$, denoted by  $A\subseteq \ma$ with $|A| = k$. Because $A$ cannot be represented as the union of multiple orbits of $G$ in $\ma$,  there exists $a\in A$ and $\sigma\in G$ such that $\sigma(a)\notin A$. It follows that $\sigma (A)\ne A$, which is a contradiction.

Next, we prove the prove the ``$\Rightarrow$'' direction by proving its contraposition: if $\orbsize{G}{\ma}$ has a sub-vector whose elements sum up to $k$, then $\fixed_G(\committee k)\ne \emptyset$.   Let $A\subseteq \ma$ denote the union of alternatives that correspond to the components of the sub-vector. Notice that for any $a\in\ma$ and any $\sigma\in G$, we have $\sigma(a) \in \orbit_G(a)$. Therefore,  $\sigma(A) = A$, which means that $G$ has a fixed point in $\committee k$.
\end{proof}

 \subsection{Proof of Claim~\ref{claim:E=Ml}}
\label{sec:proof-claim:E=Ml}

\appClaim{{\bf\boldmath $\prefspace = \committee \ell$}}{claim:E=Ml}
{For any $G\leqslant \sgroup$, any $1\le \ell\le m$,  and any partition $\vec m$ of $m$,
\begin{itemize}
\item [(i)] $ \fixed_G(\histset{m}{n}{\committee \ell})\ne\emptyset \Rightarrow n$ is feasible by  $\lcmset{\orbsize{G}{\ma}, \ell}{\oslash}$ 
\item [(ii)]  $n$ is feasible by  $\lcmset{\vec m, \ell}{\oslash}\Rightarrow \fixed_{G_{\vec m}}(\histset{m}{n}{\committee \ell})\ne\emptyset$.
\end{itemize}}

\begin{proof} {\bf Proof of part (i) of Claim~\ref{claim:E=Ml}.} 
Suppose $G$ has a fixed point in $\histset{m}{n}{\prefspace}$, denoted by $\vec h$. Let $O_1^{\committee \ell},\ldots, O_T^{\committee \ell}$ denote the orbits in ${\committee \ell}$ under $G$ such that each $\ell$-committee in each orbit appears at least once in $\vec h$, that is,
$$\forall 1\le t\le T,\forall R\in O_t^{\committee \ell}, [\vec h]_R >0$$
We note that $O_1^{\committee \ell},\ldots, O_T^{\committee \ell}$ may not be a partition of $\committee \ell$, because some $\ell$-committees may not appear in $\vec h$. Because $\vec h$ is a fixed point of $G$ in $\histset{m}{n}{\prefspace}$, the $\ell$-committees in the same obit $O_t^{\committee \ell}$ appear for the same number of times in $\vec h$. For every $1\le t\le T$, fix an arbitrary committee $A_t\in O_t^{\committee \ell}$. We have $\sum_{t=1}^T|O_t^{\committee \ell}|\times [\vec h]_{R_t} = n$. Therefore, 
$$n\in \intcomb{|O_1^{\committee \ell}|,\ldots, |O_T^{\committee \ell}|}$$
Let $O_1^\ma, O_2^\ma, \ldots, O_S^\ma$ denote the orbits in $\ma/G$ in the non-decreasing order w.r.t.~their sizes, which means that $\orbsize{G}{\ma}  = (|O_1^\ma|,\ldots, |O_S^\ma|)$. To prove $\orbsize{G}{\ma}\in \vpa{m,n}{\ell}$, it suffices to prove that for every $1\le t\le T$, there exists a partition $\vec\ell$  of $\ell$ such that  $\lcm\left(\orbsize{G}{\ma}\oslash\vec\ell\right)$ is a divisor of $|O_t^{\committee \ell}|$. We explicitly construct $\vec\ell$ as follows.   
$$\vec \ell\triangleq (|O_1^\ma\cap A_t|,\ldots, |O_S^\ma\cap A_t|)$$ 
It is not hard to verify that $\vec\ell$ is a partition of $\ell$. Next, we prove
\begin{equation}
\label{equ:lcm-div-Ot} \forall 1\le t\le T, 1\le s\le S, \frac{\lcm([\orbsize{G}{\ma}]_s,[\vec\ell]_s)}{[\vec \ell]_s}\text{ is a divisor of }|O_t^{\committee \ell}|
\end{equation}
To prove \eqref{equ:lcm-div-Ot}, we first prove the following claim. 
\begin{claim}
\label{claim:lcm-divisor}
For any $G\le \sgroup$, any $1\le \ell\le m$, any $A\in \committee \ell$, and any orbit $O\in \ma/G$ with $O\cap A\ne \emptyset$,
$$\frac{\lcm(|O|,|O\cap A|)}{|O\cap A|}\text{ is a divisor of }|\orbitset{G}{\committee \ell}(A)|$$
\end{claim}
\begin{proof}
To simplify notation, we let $A^* \triangleq O\cap A$, $\ell^* \triangleq |A^*|$, and $m^*\triangleq |O|$. Claim~\ref{equ:lcm-div-Ot} follows after the following two observations: 
\begin{equation}
\label{equ:divisor1}\frac{\lcm(m^*,\ell^*)}{\ell^*} \text{ is a divisor of } |\orbitset{G}{\committee {\ell^*}}(A^*)|,
\end{equation}
and  
\begin{equation}
\label{equ:divisor2}|\orbitset{G}{\committee {\ell^*}}(A^*)| \text{ is a divisor of } |\orbitset{G}{\committee \ell}(A)|
\end{equation}
Notice that \eqref{equ:divisor1} is equivalent to $m^*$ being a divisor of $\ell^*\times |\orbitset{G}{\committee {\ell^*}}(A^*)|$.  Also notice that $\ell^*\times |\orbitset{G}{\committee {\ell^*}}(A^*)|$ is the size of the multi-set that consists of all alternatives in all $\ell^*$-committees in the orbit of $A^*$ under $G$. That is,
$$\hat M \triangleq \bigcup \orbitset{G}{\committee {\ell^*}}(A^*)$$
The hat  on $\hat M$ indicates that it is a multi-set. Therefore, it suffices to prove that every alternative $a\in O$ appears the same number of times in $\hat M$. In fact, because $\orbitset{G}{\committee {\ell^*}}(A^*)$ is a subgroup of $G$, $G$ can be partitioned to $\frac{|G|}{|\orbitset{G}{\committee {\ell^*}}(A^*)|}$ left cosets of $\orbitset{G}{\committee {\ell^*}}(A^*)$, and every $B\in \orbitset{G}{\committee {\ell^*}}(A^*)$ is the image of a left cosets of $\orbitset{G}{\committee {\ell^*}}(A^*)$ on $A^*$. Let $\hat M^*\triangleq \bigcup_{\sigma\in G} \sigma(A^*) $  denote the multi-set that consists of all alternatives in the imagines of $A^*$ under $G$, we have
$$ \hat M^*  = |\stabset{G}{\committee {\ell^*}}(A^*)|\times  \hat M$$
Viewing $\hat M^*$ as $\bigcup_{a\in A^*} G(a)$, we note that for every $a\in A^*$, $G(a) = O$ (because $O\in \ma/G$). Therefore, $\hat M^* = |A^*|\times O$, which means that each alternative $a\in O$ appears the same number of times. It follows that each alternative $a\in O$ appears the same number of times in $\hat M$ as well, which proves \eqref{equ:divisor1}. 

Next, we prove \eqref{equ:divisor2}. Notice that for any $\sigma\in G$, if $\sigma(A)=A$, then we must have $\sigma(A^*) = A^*$, because only alternatives in $O$ can be mapped to $O$ by permutations in $G$. Therefore,  $\stabset{G}{\committee {\ell}}(A)$is a subgroup of $\stabset{G}{\committee {\ell^*}}(A^*)$. It follows from Lagrange's theorem that  $|\stabset{G}{\committee {\ell}}(A)|$ is a divisor of $|\stabset{G}{\committee {\ell^*}}(A^*)|$. Meanwhile, the orbit-stabilizer theorem (applied to $\committee {\ell^*}$ and $\committee {\ell}$) gives us
$$|G| = |\orbitset{G}{\committee {\ell^*}}(A^*)|\times |\stabset{G}{\committee {\ell^*}}(A^*)| = |\orbitset{G}{\committee {\ell}}(A )|\times |\stabset{G}{\committee {\ell}}(A )|,$$
which implies \eqref{equ:divisor2} and completes the proof of Claim~\ref{claim:lcm-divisor}.
\end{proof}
\eqref{equ:lcm-div-Ot} then follows after the applications of Claim~\ref{claim:lcm-divisor} to $O = O_s^\ma$ and $A = A_t$ for all $1\le t\le T, 1\le s\le S$.  This completes the proof of  part (i) of Claim~\ref{claim:E=Ml}.

\myparagraph{\bf Proof of part (ii) of Claim~\ref{claim:E=Ml}.} Let $\vec m = (m_1,\ldots,m_S)$ denote a partition of $m$ such that $n$ is feasible by  $\lcmset{\vec m, \ell}{\oslash}$. We will  explicitly construct a permutation group $G_{\vec m}^*$ and prove that it satisfies the desired properties. 

Suppose  $\lcm(\vec m) = \lcm (m_1,\ldots,m_S) = p_1^{q_1^{\max}} \times p_2^{q_2^{\max}} \times \cdots p_T^{q_T^{\max}}$, where $p_1,\ldots,p_T$ are different prime numbers and for every $1\le s\le S$, $q_s^{\max}\in\mathbb N$. Let $\vec p \triangleq (p_1,\ldots,p_S)$ and for every $\vec q=(q_1,\ldots,q_S)\in \mathbb Z_{\ge 0}^S$, define 
$$\vec p\,^{\vec q} \triangleq p_1^{q_1}\times p_2^{q_2}\times \cdots p_S^{q_S}$$
Using this notation, we have $\lcm(\vec m) =\vec p\,^{\vec q\,^{\max}}$, where $\vec q\,^{\max} = (q_1^{\max},\ldots,q_T^{\max})$. For any $1\le s\le S$, let 
$$m_s = \vec p^{\vec q^s} \text{ and }\gcd(d_s,d_s') = \vec p^{\vec q^{s*}},$$
where $\vec q^s = (q_1^s,\ldots,q_T^s)\in \mathbb Z_{\ge 0}^T$ and $\vec q^{s*} = (q_1^{s*},\ldots,q_T^{s*})\in \mathbb Z_{\ge 0}^T$. It follows that for all $t\le T$, $q_t^{\max} =\max_{s\le S}\{q_g^s\}$.

\begin{dfn}[{\bf\boldmath $\committee{\vec m}$}{}]
\label{dfn:A-for-G-d-star} Given $\vec m$ with $\lcm(\vec n) = p_1^{q_1^{\max}} \times p_2^{q_2^{\max}} \times \cdots p_T^{q_T^{\max}}$, define $\committee{\vec m} = \bigcup_{s=1}^S A_s \subseteq {\mathbb Z}_{\ge 0}^{q_1+\cdots + q_T +1}$, where 
\begin{align*}
A_s \triangleq  
 & \{s\}\times \prod_{t=1}^T \left(\{0,\ldots, p_t-1\}^{q_t^s}\times \{0\}^{q_t-q_t^s}\right)
\end{align*}
\end{dfn}
\begin{ex}
\label{ex:Ad}
Let $m= 14$ and $\vec m = (8,6)$. That is, $S=2$. We have $\lcm(\vec m) = 2^3\times 3 = (2,3)^{(3,1)}$, i.e., $T=2$, $p_1=2$, $q_1^{\max}=3$, $p_2=1$, $q_2^{\max}=1$. Because $m_1 =8 = 2^3\times 3^0$ and $m_2=6=2^1\times 3^1$, we have $\committee {(8,6)} = A_1\cup A_2$ defined as follows.
\begin{center}
\begin{tabular}{|r c *{4}{cc}|}
\cline{4-10}
\multicolumn{1}{c}{ } & & & \multicolumn{5}{|c|}{$p_1=2,q_1^{\max}=3$} & \multicolumn{2}{c|}{$p_2=3,q_2^{\max}=1$}\\ 
\hline 
& group & & 2 & & 2 & & 2 & & 3\\
$A_1=$ & $\{ 1\}$ & $\times$ & $\{0,1\}$ & $\times$ & $\{0,1\}$ & $\times$ & $\{0,1\}$ & $\times$ & $\{0\}$ \\
$A_2=$ & $\{ 2\}$ & $\times$ & $\{0,1\}$ & $\times$ & $\{0\}$ & $\times$ & $\{0\}$ & $\times$ & $\{0,1,2\}$ \\
\hline
\end{tabular}
\end{center}
\end{ex}
\begin{dfn}[{\bf\boldmath $G_{\vec m}^*$}{}]
\label{dfn:G-for-Al}
Given $\vec m$ with $\lcm(\vec m) = p_1^{q_1^{\max}} \times p_2^{q_2^{\max}} \times \cdots p_T^{q_T^{\max}}$, define
$$G_{\vec m}^*\triangleq \left\{\sigma_{\vec p}:\vec p\in \{0\}\times \prod_{t=1}^T\{0,\ldots,p_t-1\}^{q_t^{\max}}\right\}$$
For any $\sigma_{\vec \beta_1}, \sigma_{\vec \beta_2}\in G_{\vec m}^*$\,, we define $\sigma_{\vec \beta_1}\circ\sigma_{\vec \beta_2}\triangleq \sigma_{\vec \beta_1+\vec \beta_2 \mod \vec p\,^*}$, where $\vec p\,^* = ( S+1,  \underbrace{p_1 ,\ldots,p_1 }_{q_1^{\max}},\ldots,\underbrace{p_T ,\ldots,p_T}_{q_T^{\max}})$ and $\mod$ is the coordinate-wise modular arithmetic. 
\end{dfn}

\begin{dfn}[{\bf\boldmath $G_{\vec m}^*$ acting on $\committee{\vec m}$}{}]
For any $\sigma_{\vec p}\in G_{\vec m}^*$ and any $ \vec p\,' \in \committee{\vec m}$, we define $\sigma_{\vec p}(\vec p\,') \triangleq  \vec p\,'+\vec p\mod \vec p\,^{s*} $, where
$$\vec p\,^{s*}\triangleq \{S+1\}\times \prod_{t=1}^T \left(\{p_t-1\}^{q_t^s}\times \{0\}^{q_t-q_t^s}\right)$$
\end{dfn}
\begin{ex}
\label{ex:Gd-star}
Continuing the setting of Example~\ref{ex:Ad}, we have $\vec p\,^* = (3,2,2,2,3)$,  $\vec p\,^{1*} = (3,2,2,2,1)$, and $\vec p\,^{2*} = (3,2,1,1,3)$.
$$\sigma_{(0,1,1,1,2)} \circ \sigma_{(0,1,1,1,2)} = \sigma_{(0,2,2,2,4)\mod \vec p\,^*} = \sigma_{(0,0,0,0,1)}$$  
$$\sigma_{(0,1,1,1,2)}(1,1,0,0,0) = (1,2,1,1,2) \mod \vec p\,^{1*}  =  (1,0,1,1,0)$$
$$\sigma_{(0,1,1,1,2)}(2,1,0,0,0) = (2,2,1,1,2) \mod \vec p\,^{2*}  =  (2,0,0,0,2)$$
\end{ex}
It is not hard to verify that $|\committee{\vec m}/ G_{\vec m}^*| = \lcm(\vec d)$ and $\committee{\vec m}/ G_{\vec m}^* = \{O_1^{\committee{\vec m}},\ldots, O_S^{\committee{\vec m}}\}$, which means that $\orbsize{G_{\vec m}^*}{\ma}  = \vec m$. 

Let $\committee {\vec m, \ell}$ denote the set of all $\ell$-committees of $\committee{\vec m}$. We use the following claim to construct a fixed point of $G_{\vec m}^*$ in $\histset{m}{n}{\committee {\vec m, \ell}}$.
\begin{claim}
\label{claim:orbit-size-lcm} Let $\committee{\vec m}$ be defined as in Definition~\ref{dfn:A-for-G-d-star}. For any $\vec\ell = (\ell_1,\ldots,\ell_S)\in \mathbb Z_{\ge 0}^S$  that is a partition of  $\ell$ such that $\vec\ell\le \vec m$, there exists $A\in \committee {\vec m, \ell}$ such that for every $1\le s\le S$, $\left|A\cap O_s^{\committee{\vec m}}\right| = \ell_s$ and $\left|\orbitset{G}{\committee {\vec m, \ell}}(A)\right| = \lcm\left({\vec m}\oslash{\vec\ell}\right)$.
\end{claim}
\begin{proof}
For every $1\le s\le S$ such that $d_s'>0$, let 
$$\gcd(d_s,d_s') = p_1^{q_1^{s*}}\times p_2^{q_2^{s*}}\times \cdots \times p_T^{q_T^{s*}},$$
where $(q_1^{s*},\ldots, q_T^{s*})\in \mathbb Z_{\ge 0}^T$. 
Define
$$\vec \Delta^s \triangleq  (\Delta_1^s,\ldots,\Delta_T^s) = \vec q^s  - \vec q^{s*}\text{ and }\vec \Delta^{\max} \triangleq (\max\nolimits_{s\le S}\{\Delta_1^s\},\ldots, \max\nolimits_{s\le S}\{\Delta_T^s\}) $$
Then, we have 
$$\frac{\lcm(m_s,\ell_s)}{\ell_s} =\frac{m_s}{\gcd(m_s,\ell_s)} = \vec p\,^{\vec   \Delta^s},$$ 
which means that 
\begin{equation}
\label{equ:lcm-p-max}
\lcm(\vec m \oslash \vec\ell) = \lcm\left(\left\{\frac{m_s}{\gcd(m_s,\ell_s)}: s\le S\right\}\right) = \vec p\,^{\vec \Delta^{\max}}
\end{equation}

For every $1\le t\le T$, define $A_s^t\triangleq \{0,\ldots, p_t -1\}^{\Delta_t^s}$ to be the first $\Delta_t^s$ ``free'' coordinates for $p_t$ and define  $B_s^t$ denote the remaining coordinates for $p_t$. That is, we can represent $A_s$ as follows.
$$A_s = \{s\}\times \prod_{t=1}^T\left(\underbrace{\{0,\ldots, p_t-1\}^{\Delta_t^s}}_{A_s^t} \times \underbrace{\{0,\ldots, p_t -1\}^{q_t^{s*}}\times \{0\}^{q_t-q_t^s}}_{B_s^t}\right)$$
Fix $A_s'\subseteq \prod_{t=1}^{T}A_s^t$ to be an arbitrary set with $|A_s'| = \frac{\ell_s}{\gcd(m_s,\ell_s)}$. We define $A_s^*\subseteq A_s$ to be the extension of $A_s'$ such that the coordinates not appear in $A_s'$ take all combinations of values. Formally, 
$$A_s^*\triangleq  \left\{\{s\}\times \prod\nolimits_{t=1}^T\left( \vec a|_{A_s^t} , \vec b\right):\vec a\in A_s', \vec b\in B_s^t\right\},$$
where $\vec a|_{A_s^t}$ is the ${A_s^t}$ components of $\vec a$.
Then we define $A \triangleq \bigcup_{s=1}^SA_s^*$ and define the following set of permutations in $G_{\vec m}^*$. 
$$G\triangleq \left\{\sigma_{\vec p}: \vec p\in \{0\}\times \prod_{t=1}^T\left( \{0 \}^{\Delta_t^{\max}}  \times \{0,\ldots, p_t -1\}^{q_t-\Delta_t^{\max}}\right)\right\}$$
\begin{ex}
Continuing the setting of Example~\ref{ex:Gd-star}, we let $\vecp d = (6,4)$, which means that $\gcd(d_1,d_1')=2 = 2^1\times 3^0$ and $\gcd(d_2,d_2')=2 = 2^1\times 3^0$. Therefore,
\begin{align*}
\vec \Delta^1 = & (3,0) - (1,0) = (2,0)\text{, }\vec \Delta^2 = (1,1) - (1,0) = (0,1)\text{, and }\vec \Delta^{\max} = (2,1)\\
A_1 = & \{1\}\times \underbrace{\{0,1\}\times \{0,1\}}_{A_1^1}\times \underbrace{\{0,1\}}_{B_1^1}\times \underbrace{\{0\}}_{B_1^2}\\
A_2 = & \{2\}\times \underbrace{\{0,1\}\times \{0\}\times \{0\}}_{B_1^1}\times \underbrace{\{0,1,2\}}_{A_1^2}
\end{align*}
Let $A_1' = \{(0,0),(0,1),(1,0)\}$ and $A_2' = \{0,1\}$. We have
\begin{align*}
A_1^* = & \{1\}\times  \{(0,0),(0,1),(1,0)\} \times \{0,1\} \times \{0\} \\
A_2^* = & \{2\}\times  \{0,1\}\times \{0\}\times \{0\} \times  \{0,1\} \\
G =& \left\{\sigma_{\vec p}: \vec p\in \{0\}\times \{0\}\times\{0\}\times\{0,1\}\times\{0\}  \right\}
\end{align*}
\end{ex}

It follows that for every $\sigma\in G$ and every $s\le S$, we have $\sigma(A_s^*) = A_s^*$. Therefore, $\sigma$ is a stabilizer of $A$ in $G_{\vec m}^*$\,, which means that  
$$\left|\stabset{G_{\vec m}^*}{\committee {\vec m, \ell}}(A)\right|\ge |G| =\prod_{t=1}^T p_t^{q_t-\Delta_t^{\max}}= \frac{\vec p\,^{\vec q\,^{\max}}}{\vec p\,^{\vec \Delta^{\max}}} = \frac{|G_{\vec m}^*|}{\lcm\left(\left\{\frac{\lcm(m_s,\ell_s)}{\ell_s}: s\le S\right\}\right)}$$
The last equation follows after \eqref{equ:lcm-p-max}. Therefore, following the orbit-stabilizer theorem, we have
\begin{equation}
\label{equ:cal-orbit-size}
\left|\orbitset{G_{\vec m}^*}{\committee {\vec m, \ell}}(A)\right| = \frac{|G_{\vec m}^*|}{\left|\stabset{G_{\vec m}^*}{\committee {\vec m, \ell}}(A)\right|}\le \lcm\left(\vec m\oslash\vec\ell\right)
\end{equation}
Claim~\ref{claim:orbit-size-lcm} follows after \eqref{equ:cal-orbit-size} and Claim~\ref{claim:lcm-divisor}.
\end{proof}
Back to the proof for the second part of Claim~\ref{claim:E=Ml}, suppose $n= \sum_{w=1}^Wn_w\times \lcm\left( {\vec m}\oslash{\vec \ell\,^w}\right)$. Let $A^1,\ldots,A^W\in\committee {\vec m, \ell}$ denote the $\ell$-committees guaranteed by Claim~\ref{claim:orbit-size-lcm} applied to $\committee{\vec m}$ (Definition~\ref{dfn:A-for-G-d-star}), $G_{\vec m}^*$ (Definition~\ref{dfn:G-for-Al}), and $\vec \ell\,^1,\ldots, \vec \ell\,^W$, respectively. Let 
$$P_{\vec d}\triangleq \bigcup\nolimits_{w=1}^W n_w\times \orbitset{G_{\vec m}^*}{\committee {\vec m, \ell}}(A^w)$$
Because for every $w\le W$ and every $\sigma\in G_{\vec m}^*$, $\sigma\left(\orbitset{G_{\vec m}^*}{\committee {\vec m, \ell}}\right) = \orbitset{G_{\vec m}^*}{\committee {\vec m, \ell}}$, we have $\hist(\sigma(P_{\vec d}))=\hist(P_{\vec d})$, which means that $\hist(\sigma(P_{\vec d})) \in \histset{m}{n}{\committee{\vec m}}$, which proves part (ii) of Claim~\ref{claim:E=Ml}.
\end{proof}
  
\subsection{Corollary~\ref{coro:ell=k=1} and Theorem~\ref{thm:at-large-ANR}}
\label{sec:proof-thm:at-large-ANR}
We first prove a useful corollary of Theorem~\ref{thm:ANR-common} about the special case of $(\prefspace,\decspace) = (\listset{1},\committee{1})$. 

\begin{coro}[{\bf\boldmath $(\prefspace,\decspace) = (\listset{1},\committee{1})$}{}]
\label{coro:ell=k=1}
For $(\listset{1},\committee{1})$ rules, the strong ANR impossibility never holds; at-large ANR impossibility holds if and only if $m=5$ or $m\ge 7$.
\end{coro}
\begin{proof} Suppose for the sake of contradiction that the strong ANR impossibility holds for some $m\ge 2$. 
By Theorem~\ref{thm:ANR-common},  there exists  a partition $\vec m$ of $m$ that does not contain $1$ and same-length partition $\vec\ell$ of $1$ such that $\vec m\circledast \vec \ell =1$. This means that the $\vec m$ component that corresponds to the $1$  component in $\vec\ell$ is $1$, which is a contradiction.

Due to basic number theory, a set of denominations $\calC$ can make any sufficiently large $n$ if and only if $\lcm(\calC) =1$. Therefore, to prove the ``if'' part of at-large ANR impossibility, it suffices to construct $\vec m$ such that $\lcmset{\vec m,1}{\circledast}$ contains co-prime numbers as follows.
\begin{itemize}
\item When $m=5$, let $\vec m = (3,2)$ and consider $\vec \ell \in \{(1,0), (0,1)\}$. Then, $\{2,3\}\subseteq\lcmset{\vec m,1}{\circledast}$.
\item When $m\ge 7$, let $\vec m$ to be defined as in  define \eqref{eq:vecd}. It is not hard to verify that  $m_1$ and $m_2$ are co-primes. Consider $\vec \ell \in \{(1,0), (0,1)\}$. Then, $\{m_1,m_2\}\subseteq\lcmset{\vec m,1}{\circledast}$.
\end{itemize}

The ``only if'' part of at-large ANR impossibility is proved in the following cases.
\begin{itemize}
\item When $m=2$, $\vda{m}{1} = \{(2)\}$. We have $\lcmset{(2),1}{\circledast} = \{2\}$, which means that the ANR impossibility does not hold for all odd $n$'s.
\item When $m=3$, $\vda{m}{1} = \{(3)\}$. We have $\lcmset{(3),1}{\circledast} = \{3\}$, which means that the ANR impossibility does not hold for all $n$'s that are not divisible by $3$.
\item When $m=4$, $\vda{m}{1} = \{(4),(2,2)\}$. We have $\lcmset{(4),1}{\circledast} = \{4\}$ and $\lcmset{(2,2),1}{\circledast} = \{2\}$, which means that the ANR impossibility does not hold for all odd $n$'s.
\item When $m=6$, $\vda{m}{1} = \{(6),(4,2), (3,3), (2,2,2\}$. We have  $\lcmset{(6),1}{\circledast} = \{6\}$, $\lcmset{(4,2),1}{\circledast} = \{4,2\}$, $\lcmset{(3,3),1}{\circledast} = \{3\}$, and $\lcmset{(2,2,2),1}{\circledast} = \{2\}$, which means that the ANR impossibility does not hold for all  $n$'s that are divisible by neither $2$ or $3$.
\end{itemize}
\end{proof}

\begin{thm}
\label{thm:at-large-ANR}  Under the \commonsettings{}, at-large ANR impossibility holds if
\begin{center}
\renewcommand{\arraystretch}{1}
\begin{tabular}{|c|*{2}{c|}}
 \hline
 \diagbox{$\prefspace$} {$\calD$} & $\listset k$ & $\committee k$ \\
 \hline
 $\listset \ell$ & $m\ge 8$,  $ \ell\le \frac m 2 -2 $,  and $ k\le m$ &  $m\ge 12$,  $\ell\le \frac m 4 -2 $, and $ k\le m-1$  \\
\hline
$\committee \ell$ &$m\ge 4$, $ \ell\le m-2$, and $ k\le m$ & 
\begin{tabular}{@{}c@{}}
$m\ge 2$, $k\le m-1$, and \\
($\ell \notin\{k,m-k\}$ or $\ell \le m/2-3$)
\end{tabular} \\ 
\hline
 \end{tabular}
 \renewcommand{\arraystretch}{1}
\end{center}
Moreover, under the conditions in the table, the ANR impossibility holds for every $n\ge m^2/2$.
\end{thm}


\begin{proof}
According to basic number theory, a set of denominations $\calC$ can make any sufficiently large $n$ if and only if $\gcd(\calC) =1$. To provide sufficient conditions for at-large ANR impossibility, we will define $\vec m$ such that $\lcmset{\vec m,\ell}{\circledast}$ contains co-prime numbers. Defined
\begin{equation*} 
\vec m = (m_1,m_2)\triangleq \begin{cases}
\left(\frac{m+1}{2},\frac{m-1}{2}\right)& 2\nmid m\\
\left(\frac{m}{2}+1,\frac{m}{2}-1\right)& 4\mid m\\
\left(\frac{m}{2}+2,\frac{m}{2}-2\right)&4\nmid  m \text{ and }  2\mid m
\end{cases}\hspace{10mm}
\eqref{eq:vecd}
\end{equation*}
It is not hard to verify that  $m_1$ and $m_2$ are co-primes and for every $n\ge m^2/2$, $n$ is feasible by $\{m_1, m_2\}$.

\myparagraph{Proof for $(\listset \ell,\listset k)$.} When $m\ge 8$, $\vec m\ge (2,2)$, which means that it satisfies the sub-vector constraint for $\decspace = \listset k$. Consider $\vec \ell \in \{ (\ell ,0), (0,\ell)\}$. When $\ell\le \frac m2-2$, $\vec \ell\le \vec m$. It follows that for every $n\ge m^2/2$, $n$ is feasible by $\{m_1, m_2\}\subseteq {\lcmset{\vec d,\ell}{\circledast}}$. The ANR impossibility holds due to the $(\listset{\ell},\committee{k})$ case of Theorem~\ref{thm:ANR-common}. 

\myparagraph{Proof for $(\listset \ell,\committee k)$.}  Consider $\vec m =(m_1,m_2)$  defined in \eqref{eq:vecd}. When $k\not\in \{m_1,m_2\}$, then  we have $\left\{m_1,m_2\right\}\subseteq \lcmset{\vec d, \ell}{\circledast} $, via $\vec\ell = (\ell ,0)$ and $ (0,\ell)$, respectively.   
When $k\in \{m_1,m_2\}$, then we consider $\vec m = (m-2\ell-1, \ell+1,\ell)$ and $\vec\ell\in \{(0,\ell,0),(0,0,\ell)\}$. It follows that  $\left\{\ell+1,\ell\right\}\subseteq \lcmset{\vec d, \ell}{\oslash}$. The ANR impossibility holds due to   the $(\listset{\ell},\committee{k})$ case of Theorem~\ref{thm:ANR-common}.

\myparagraph{Proof for $(\committee \ell,\listset k)$.}  Let $\vec m = (\ell,m-\ell)$ and $\vec\ell = (\ell, 0)$. Then $1\in \lcmset{\vec d, \ell}{\oslash}$. The ANR impossibility holds due to   the $(\committee \ell,\listset k)$ case of Theorem~\ref{thm:ANR-common}.

\myparagraph{Proof for $ (\committee{\ell},\committee{k})$.}  When $\ell\notin\{k,m-k\}$, then we let $\vec m = (\ell,m-\ell)$ and $\vec\ell  = (\ell,0)$, which means that $1\in \lcmset{\vec d,\ell}{\oslash}$. Therefore, any $n$ is feasible. Because $\ell\notin\{k,m-k\}$, the sub-vector constraint is satisfied. When $\ell\in\{k,m-k\}$ and $\ell \le m/2-3$, define $\vec m$ as in \eqref{eq:vecd}.  The ANR impossibility holds due to   the $(\committee \ell,\committee k)$ case of Theorem~\ref{thm:ANR-common}.
\end{proof}

\begin{lem}
\label{lem:relationship}
For any $m$, $n$, $1\le \ell^*\le \ell\le m$, and $1\le k^*\le k\le m$, we have the following relationship between the ANR impossibilities for different combinations of $(\prefspace,\calD)$, where an edge $(\prefspace,\decspace)\ra (\prefspace',\decspace')$ mean that if the ANR impossibility holds for the source   setting $(\prefspace,\decspace)$, then it holds for the sink setting $(\prefspace',\decspace')$ as well.
\begin{figure}[htp]
\centering
\includegraphics[width = .7\textwidth]{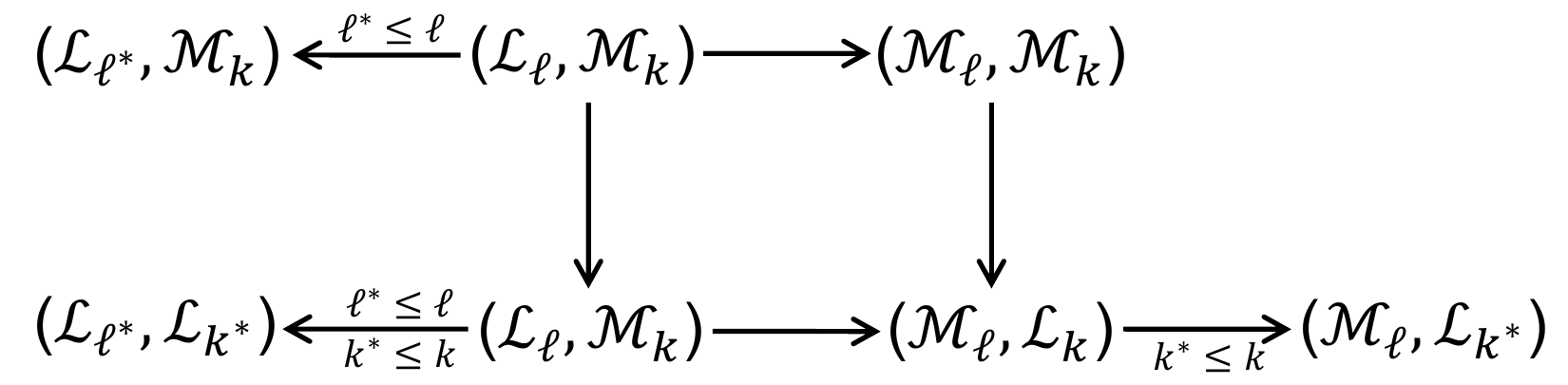}
\end{figure}
\end{lem}
\begin{proof} {$(\listset{\ell},\committee{k})\ra (\committee{\ell},\committee{k})$ and $(\listset{\ell},\listset{k})\ra (\committee{\ell},\listset{k})$:} For any $\vec d$ and $\vecp d$, we have $\lcm(\vec d\oslash \vecp d) \mid \lcm(\vec d\circledast \vecp d) $. Therefore, $\vpl{m,n}{\ell}\subseteq \vpa{m,n}{\ell}$.

{$(\listset{\ell},\committee{k})\ra (\listset{\ell},\listset{k})$ and $(\committee{\ell},\committee{k})\ra (\committee{\ell},\listset{k})$:} This follows after $\vdl{m}{k}\subseteq \vda{m}{k}$.

{$ (\listset{\ell},\committee{k})\ra (\listset{\ell'},\committee{k})$, $ (\listset{\ell},\listset{k})\ra (\listset{\ell'},\listset{k'})$, and $ (\committee{\ell},\listset{k})\ra (\committee{\ell},\listset{k'})$:} This follows after (1) $\vdl{m}{k}\subseteq \vdl{m}{k'}$, because any $\vec d$ with no more than $k$ ones has no more than $k'\ge k$ ones;  and (ii) $\vpl{m,n}{\ell}\subseteq \vpl{m,n}{\ell'}$, because for any $\vecp d\in \simplex{\ell}{0}$, there exists $\vec d\,''\in \simplex{\ell}{0}$ with $\vec d\,''\le \vecp d$.
\end{proof}

\subsection{Proof of Theorem~\ref{thm:at-large-up-to-L}}
\label{sec:proof-at-large-up-to-L}

\appThm{\bf At-large ANR impossibility: up-to-$L$ preferences}{thm:at-large-up-to-L}
{
For any $(\prefspace,\decspace)\in \{\listset{\le L}, \committee{\le L}:1\le L\le m\}\times \{\listset{k}, \committee{k}:1\le k\le m\}$, at-large ANR impossibility holds if and only if 
\begin{center}
\renewcommand{\arraystretch}{1}
\begin{tabular}{|c|*{2}{l|}}
 \hline
 \diagbox{$\prefspace$} {$\calD$} & \hspace{23mm}$\listset k$ & \hspace{23mm}$\committee k$ \\
 \hline
 $\listset {\le L}$ &
 \begin{tabular}{@{}l@{}} (i) $m=5$, or \\
 (ii) $m\ge 7$, or \\(iii) $k \ge 2$  \end{tabular} 
 &  
 \begin{tabular}{@{}l@{}} (i) $m=5$ and $k\le m-1$, or \\
 (ii) $m\ge 7$ and $k\le m-1$, or \\(iii) $2\le k\le m-2$  \end{tabular}\\
\hline
$\committee {\le L}$
&
\begin{tabular}{@{}l@{}}  
 (i) $m=5$, or\\
 (ii) $m\ge 7$, or \\
 (iii) $\max(L,k)\ge 2$,  \\
 except $(m,L,k)\in\{ (2,1,2),  (3,2,1)\}$ 
 \end{tabular} & 
\begin{tabular}{@{}l@{}}
 (i) $m=5$ and $k\le m-1$, or\\
  (ii) $m\ge 7$ and $k\le m-1$, or\\
 (iii) $[L]\not\subseteq \{k,m-k\}$ and $k\le m-1$\\
 \ \\
\end{tabular} \\ 
\hline
 \end{tabular}
 \renewcommand{\arraystretch}{1}
\end{center}
Moreover, under the conditions in the table, the ANR impossibility holds for every $n\ge m^2/2$.
}

\begin{proof} The proof is done by applications of Theorem~\ref{thm:ANR-common} to all cases. Following a similar idea in the proof of Theorem~\ref{thm:at-large-ANR}, to prove that at-large ANR impossibility holds, we specify a partition $\vec m$ of $m$ (that satisfies the sub-vector constraint) and two  $\vec \ell$'s so that the two coins created by them are co-primes.

\myparagraph{Proof for $(\listset {\le L},\listset k)$.}  
We first prove that when $\prefspace = \listset {\le L}$, it suffices to focus on $L=1$ case.
\begin{claim}
\label{claim:suffice-ell=1}
For any $m,n,k$ and $L$, the ANR impossibility holds for $(\listset{\le L},\listset{k})$ (respectively, $(\listset{\le L},\committee{k})$)  if and only if it holds for $(\listset{1},\listset{k})$ (respectively, $(\listset{1},\committee{k})$).
\end{claim}
\begin{proof}
Due to Lemma~\ref{lem:ANR-union}, it suffices to prove that for any permutation group $G$, the coins created by $\listset{1}$, i.e., $\orbsize{\listset{1}}{G}$, are finer than the coins created by any $\listset{1}$, i.e., $\orbsize{G}{\listset{j}}$. To see this, let $a$ denote the top-ranked alternative in $R$ and consider the applications of the orbit-stabilizer theorem to $G$ on $\listset{j}$ and on $\listset{1}$, respectively: 
\begin{align*}
G\text{ on }\listset{j}: & |G| = |G(R)|\times |\stab_{G}(R)|\\
G\text{ on }\listset{1}: & |G| = |G(a)|\times |\stab_{G}(a)|
\end{align*}
Therefore, 
$$|G(R)|\times |\stab_{G}(R)| = |G(a)|\times |\stab_{G}(a)|$$
Notice that any stabilizer $g\in \stab_{G}(R)$, which means that $g(R) =R$, preserves the top-ranked alternative in $R$, which means that $g$ is also a stabilizer of $a$. Therefore, $ \stab_{G}(a)$ is a subgroup of $\stab_{G}(R)$. It follows from Lagrange's theorem that $|\stab_{G}(R)|$ divides $|\stab_{G}(a)|$. Therefore, $|G(a)|$ divides $|G(R)|$, which means that $|G(a)|$ is a finer coin in $\orbsize{\listset{1}}{G}$. This proves the claim. 
\end{proof}

By Claim~\ref{claim:suffice-ell=1}, it suffices to characterize the ANR impossibility for  $(\listset{1},\listset{k})$. The ``if'' direction is proved in the following cases.  When $k\ge 2$, let $\vec m = (m-1,1)$, and $\ell = (0,1)$. Then $1\in \lcmset{\vec m,1}{\circledast}$. By Theorem~\ref{thm:ANR-common}, the ANR impossibility holds for all $n\ge 1$. When $m=5$ and $k=1$, let $\vec m = (3,2)$, $\vec \ell\in \{(0,1),(1,0)\}$. Then $\{2,3\} \subseteq \lcmset{\vec m,1}{\circledast}$.  When $m\ge 7$ and $k=1$, let $\vec m$ be defined as in \eqref{eq:vecd} and let $\ell\in \{(0,1),(1,0)\}$. Then $\lcmset{\vec m,1}{\circledast}$ contains two co-prime numbers. 

The ``only if'' direction if proved by enumerating all $\vec m$ that satisfies the sub-vector constraint (for $k=1$)  and all $n$'s for which the ANR impossibility does not hold as summarized in the following table.  
\begin{center}
\renewcommand{\arraystretch}{1.5}
\begin{tabular}{|r | *{2}{c|}}
\hline 
$m$ & partitions satisfying the sub-vector constraint  &  ANR Imp does not hold for\\
\hline 
$6$ & $\{(6), (4,2), (3,3),(2,2,2)\}$ & $2\nmid n$ and $3\nmid n$\\
\hline 
$4$ & $\{(4), (2,2)\}$ & $2\nmid n$\\
\hline $3$ & $\{(3) \}$ & $3\nmid n$\\
\hline $2$ & $\{(2) \}$ & $2\nmid n$\\
\hline
\end{tabular}
\renewcommand{\arraystretch}{1}
\end{center}

\myparagraph{Proof for $(\listset {\le L},\committee k)$.}  Due to Claim~\ref{claim:suffice-ell=1}, the rest of the proof focuses on characterizing ANR impossibility for  $(\listset{1},\committee{k})$.

The ``if'' direction. 
\begin{itemize}
\item When $m=5$ and $k\in \{1,m-1\}$, let $\vec m = (m-1,1)$, $\vec \ell = (0,1)$. Then $\{2,3\} \subseteq \lcmset{\vec m,1}{\circledast}$. 
\item When $m\ge 7$ and $k\in \{1,m-1\}$, let $\vec m$ be defined as in \eqref{eq:vecd} and let $\vec \ell\in \{(0,1),(1,0)\}$. Then $\lcmset{\vec m,1}{\circledast}$ contains two co-prime numbers. 
\item When $2\le k\le m-2$, let $\vec m = (m-1,1)$,  $\vec \ell = (0,1)$. Then $1\in \lcmset{\vec m,1}{\circledast}$. This means that the ANR impossibility holds for all $n\ge 1$.
\end{itemize}

The ``only if'' direction. 
\begin{itemize}
\item When $k=m$, any partition $\vec m$ of $m$ sum up to $k=m$, which means that the ANR impossibility does not hold as $\vda{m}{k}=\emptyset$. Recall that $\vda{m}{k}$ is the set of all partitions of $m$ that satisfies the sub-vector constraint for $\decspace = \committee k$ in Theorem~\ref{thm:ANR-common} as defined in  Definition~\ref{dfn:partitions}. 
\item When $m=6$ and $k\in \{1,m-1\}$, we have $\vda{m}{k} = \{(6), (4,2), (3,3),(2,2,2)\}$. It follows that for every $n$ such that $2\nmid n$ and $3\nmid n$, the ANR impossibility does not hold. 
\item When $m=4$ and $k\in \{1,m-1\}$, we have $\vda{m}{k} = \{(4), (2,2)\}$. Therefore, the ANR impossibility does not hold for every $n$ with $2\nmid n$.
\item When $m=3$ and $k\in \{1,m-1\}$, we have $\vda{m}{k} = \{(3)\}$. Therefore, the ANR impossibility does not hold for every $n$ with $3\nmid n$.
\item When $m=2$ and $k\in \{1,m-1\}$, we have $\vda{m}{k} = \{(2)\}$. Therefore, the ANR impossibility does not hold for every $n$ with $2\nmid n$.
\end{itemize}

\myparagraph{Proof for $(\committee {\le L},\listset k)$.}  The ``if'' direction.   
\begin{itemize}
\item When $\max(L,k)\ge 2$ and $m\ge 4$, there are two sub-cases. If $L \ge 2$, then consider the partition $\vec m = (m-2,2)$, which does not contain $1$. Therefore, $\vec m\in \vdl{m}{k}$. In light of Lemma~\ref{lem:ANR-union}, it suffices to show that at-large ANR impossibility holds for $\ell =2\le L$. Let $\vec\ell = (0,2)$. We have $\lcm(\vec m\oslash \vec\ell ) =1$, which means at-large ANR impossibility holds. If $k \ge 2$, then consider $\vec m = (m-1,1)$, $\ell =1$,  and $\vec\ell = (0,1)$, which means that  $\lcm(\vec m\oslash \vec\ell ) =1$ and therefore at-large ANR impossibility holds.
\item When $\max(L,k)\ge 2$, $m=3$,  and $(L,k) \ne (2,1)$, we prove the theorem by explicitly constructing $\vec m$ and $\vec
\ell$ in the following two sub-cases. If $k\ge 2$, then we let $\vec m = (2,1)$, $\ell=1$,  and $\vec\ell = (0,1)$. Otherwise $k=1$ and $L\ge 3$, which means that $L=3=m$. In this case we let $\vec m = (3)$ and $\vec\ell = (3)$.
\item When $\max(L,k)\ge 2$, $m=2$, and $(\ell,k)\ne (1,2)$, we have $\ell =2$. The theorem is proved by letting $\vec m = (2)$ and $\vec\ell = (2)$. 
\item When $\max(L,k)=1$ and either $m=5$ or $m\ge 7$, at-large ANR impossibility holds according to Corollary~\ref{coro:ell=k=1}.
\end{itemize}
The ``only if'' direction.   
\begin{itemize}
\item When $(m,L,k)\in\{ (2,1,2),  (3,2,1)\}$,  we have $\vda{m}{k} =\{(m)\}$, which means that at-large ANR impossibility does not hold.
\item When $\max(L,k)=1$ and $m\in \{2,3,4,6\}$, at-large ANR impossibility does not hold according to Corollary~\ref{coro:ell=k=1}.
\end{itemize}

\myparagraph{Proof for $(\committee {\le L},\committee k)$.}   When $[L]\not\subseteq \{k,m-k\}$, there exists $\ell \le L$ such that $\ell\notin  \{k,m-k\}$. Then, at-large ANR impossibility holds by  letting $\vec m = (m-\ell,\ell)$ and $\vec \ell = (0,\ell)$ in Theorem~\ref{thm:ANR-common}.   When $[L]\subseteq \{k,m-k\}$. There are two cases:
\begin{itemize}
\item Case 1: $L =1$ and $k=1$ or $m-1$. The theorem follows after Corollary~\ref{coro:ell=k=1}: at-large ANR impossibility holds if and only if $m=5$ or $m\ge 7$. 
\item Case 2: $L =2$, $m=3$, and $k=1$ or $2$. In this case $\vda{3}{k} = \{(3)\}$, which means that $\lcmset{{3},2}{\circledast} = \{3\}$. Or equivalently, at-large ANR impossibility does not hold.
\end{itemize}

\end{proof}



\section{Materials for Section~\ref{sec:alg}}
\label{app:alg}


\subsection{Proof of Theorem~\ref{thm:MFP}}
\label{sec:proof-thm:MFP}
\appThmnoname{thm:MFP} {
For any polynomially computable $f$ and any $(\listset m,\decspace)$ in the \commonsettings{}, Algorithm~\ref{alg:tie-breaking}  computes $\hpp_\tb$ in polynomial time.
}
\begin{proof} We first prove that the choice of $\sigma$ in the definition of $\hpp$ does not matter. More precisely, we prove that for any profile $P$, any $\sigma_1,\sigma_2\in \hp(P)$, and any fixed point $d\in \fixed_{\stab(\hist(P))}(\calD)$, we have $\sigma_1(d)=\sigma_2(d)$. Suppose for the sake of contradiction that $\sigma_1(d)\ne \sigma_2(d)$. Notice that $\hist(\sigma_1(P))=\hist(\sigma_2(P))$. Therefore, $\hist(\sigma_2^{-1}\circ\sigma_1(P)) = \hist(\sigma_2^{-1}\circ\sigma_2(P)) = \hist(P)$, which means that $\sigma_2^{-1}\circ\sigma_1\in\stab(\hist(P))$. Because $\sigma_2^{-1}\circ\sigma_1(d) \ne d$, $d$ is not a fixed point of $\stab(\hist(P))$, which is a contradiction.

We prove that $\breaking{\hpp_\tb}{\cor}$ is a most equitable refinement of $\cor$ by proving that for every $P\notin \problematic{\cor}{n}{\prefspace}{\decspace}$, $\anr(\hpp_f\ast\cor,P)=1$.  Intuitively, this is true because any MFP does not depend on the identity of agents or the decisions. Formally, we have the following proof.

\myparagraph{$\hpp_f\ast\cor$ satisfies anonymity at $P$.} It suffices to prove that for any profile $P'$ with $\hist(P')=\hist(P)$, $\hp(P)=\hp(P')$. This follows after noticing that for any permutation $\sigma\in\sgroup$,
$$\hist(\sigma(P)) =\sigma(\hist(P))=\sigma(\hist(P')) = \hist(\sigma(P'))$$
More precisely, any $\sigma\in\sgroup$ that maximizes $\hist(\sigma(P))$ according to $\rhd$ would also maximize $\hist(\sigma(P'))$. 

\myparagraph{$\hpp_f\ast\cor$ satisfies neutrality at $P$.} Suppose for the sake of contradiction that $\hpp_f\ast\cor(P) = \{a\}$ and there exists a permutation $\sigma\in\sgroup$ such that $\hpp_f\ast\cor(\sigma(P)) = \{b\}$, where $b\ne \sigma(a)$. Let $\sigma_a\in \hp(P)$ and $\sigma_b\in \hp(\sigma(P))$ denote any pair of permutations. We show that $P$ and $\sigma(P)$ are ``similar'' by proving the following two properties.

\begin{itemize}
\item  [(i)] $\hist(\sigma_a(P))=\hist(\sigma_b(\sigma(P)))$.  Suppose for the sake of contradiction that  this does not hold and w.l.o.g.~$\hist(\sigma_a(P))\rhd\hist(\sigma_b(\sigma(P)))$. Then, notice that $\hist(\sigma_a(P)) = \hist(\sigma_a\circ \sigma^{-1}(\sigma(P)))$, which means that $\sigma_a\circ \sigma^{-1}$ maps $\hist(\sigma(P))$ to a profile that is ranked higher than $\sigma_b(\hist(\sigma(P)))$. This contradicts  the optimality of $\sigma_b$, i.e., $\sigma_b\in \hp(P)$.
\item[(ii)] $\sigma(a)\in \fw(\sigma(P),\sigma(\cor(P)))$. Suppose for the sake of contradiction that $\sigma(a)\notin \fw(\sigma(P),\sigma(\cor(P)))$. Because $\sigma(a)\in\sigma(\cor(P))$, we must have that $\sigma(a)$ is a fixed point under $\stab(\hist(\sigma(P))$, which means that there exists $\sigma'\in \stab(\hist(\sigma(P)))$ such that  $\hist(\sigma'(\sigma(P))) = \hist(\sigma(P))$ and $\sigma'(\sigma(a))\ne \sigma(a)$. Let $\sigma^* = \sigma^{-1}\circ \sigma'\circ\sigma$, we have $\hist(\sigma^*(P)) =\hist( P)$ and $\sigma^*(a)\ne a$, which means that $a$ is not a fixed point under $\stab(\hist(P))$, which is a contradiction. 
\end{itemize}
It follows from (i) that $\sigma_b\circ\sigma(P)$ is the most-preferred histogram (which is the same as $\hist(\sigma_a(P))$ among permuted histograms according to $\rhd$ defined in Defininition~\ref{dfn:ext-priority}. Therefore, $\sigma_b\circ\sigma \in \hp(P)$. Notice that because $b\in \fw(\sigma(P),\sigma(\cor(P)))$, we have $\sigma^{-1}(b)\in \fw(P)\cap\cor(P)$ according to (ii), where we switch the roles of $a$ and $b$. Also recall that $b\ne \sigma(a)$. Therefore, $\sigma^{-1}(b)\ne a$. Because of the optimality of $a$ and Proposition~\ref{prop:hpp}, $a$ has higher priority than $\sigma^{-1}(b)$ for any permutation in $\hp(P)$, especially $\sigma_b\circ\sigma$.  Therefore,  
$$\sigma_b(\sigma(a))= \sigma_b\circ\sigma(a)\rhd \sigma_b\circ\sigma(\sigma^{-1}(b)) = \sigma_b(b),$$
which contradicts the optimality of $b$, as $\sigma(a)\in\fw(\sigma(P),\sigma(\cor(P)))$ according to (ii). 

\myparagraph{$\hpp_f\ast\cor$ is resolute at $P$.} This part follows after the definition of $\hpp_f\ast\cor$. 
\end{proof}

\subsection{Proof of Theorem~\ref{thm:hpp-opt}}
\label{sec:proof-thm:hpp-opt}
\appThmnoname{thm:hpp-opt}{
For any polynomially computable $f$ and under the \commonsettings{} where $\prefspace = \listset m$, Algorithm~\ref{alg:tie-breaking} runs in polynomial time and computes $\hpp_\tb$.
}
%
\begin{proof}  We first verify that Algorithm~\ref{alg:tie-breaking} correctly computes an $\hpp$ breaking. \eqref{equ:unc-comp} holds because $\sigma\in\stab(\hist(P))$ if and only if for every ranking $R$, $[\hist(P)]_R = [\hist(P)]_{\sigma(R)}$. Let $R^* = \arg\max_{R\in \text{MPR}(P)}^\rhd  \hist(\sigma_R(P))$. To verify that $\sigma_{R^*}$   is indeed a highest-priority permutation, for the sake of contradiction suppose there exists $\sigma\in\sgroup$ such that $\hist(\sigma(P))\rhd \hist(\sigma_{R^*}(P))$. This means that  $[\hist(\sigma(P))]_\rhd\ge [\hist(\sigma_{R^*}(P))]_\rhd >0$, where $[\hist(\sigma(P))]_\rhd$ is the $\rhd$ coordinate of $\hist(\sigma(P))$, or in other words, the multiplicity of ranking $\rhd = [1\succ\cdots\succ m]$ in $\sigma(P)$. Therefore, $\sigma^{-1}(\rhd)$ must be a most popular ranking in $P$. This contradicts the maximality of ${R^*}$.

Next, we verify that Algorithm~\ref{alg:tie-breaking}  runs in polynomial time in $m$, $n$, and $|D|$. \eqref{equ:unc-comp} takes $O(mn^3)$ time. Step~\ref{step:fixed} takes $\text{poly}(mn)|D|$ time, because it only need to verify whether every $d\in D$ is a fixed point of $\stab(\hist(P))$. In step~\ref{step:sigmastar}, computing  $\hist(\sigma_R(P))$ (in the list form) for each $R$ takes $O(mn + m\log n)$ time, and each comparison when computing the $\arg\max_{ R\in \text{MPR}} ^\rhd$ takes $\text{poly}(nm)$ time, which means that the overall time for step~\ref{step:sigmastar} is $\text{poly}(nm)$. Step~\ref{step:return} takes $\text{poly}(m|D|)$ time. 
\end{proof}
 
\section{General Settings}
\label{sec:general}


\subsection{Definitions of Anonymity And Neutrality For General Settings}
\label{sec:dfn-general}
{\bf Intuition.} The  anonymity   for  voting rule  $\cor:\prefspace^n\ra\decspace$ in the general setting is straightforward. To define neutrality, we need a sensible and consistent way to capture the following idea behind the neutrality:
\begin{center}
\em when agent permutes their preferences in a certain way,   the winner  is permuted in the same way
\end{center}
This is achieved by leveraging any permutation $\sigma$ over $\ma$ to a permutation over $\prefspace$ and a permutation over $\decspace$. Specifically, for any pair  of permutations over $\ma$, $\sigma_1$ and $\sigma_2$, first applying the counterpart of $\sigma_1$ to $\prefspace$ and then applying the counterpart of $\sigma_2$ should be the same as directly applying the counterpart of $\sigma_2\circ\sigma_1$ (which is a permutation over $\ma$). Such consistency should be enforced for $\decspace$ as well. 

This idea is captured in a well-studied notion in group theory called {\em group actions}. 
Basic definitions and notation about group theory can be found in Appendix~\ref{app:prelim}.

\begin{dfn}[{\bf Group actions}{}]
\label{dfn:group-action}
A group $G$ {\em acts} on a set $X$, if every $g\in G$ can be viewed as a permutation on $X$, such that (1) for all $g_1,g_2\in G$ and all $x\in X$, we have $g_1(g_2(x)) = g_1\circ g_2(x)$, where $\circ$ is the operation in $G$, and (2) let $\id\in G$ denote the identity, then for all $x\in X$, we have $\id(x) =x$.
\end{dfn}
The permutations defined on $k$-committees, $k$-lists,  profiles, and histograms in Section~\ref{sec:prelim} are examples of $\sgroup$ acting on   $k$-committees, $k$-lists,  profiles, and histograms, respectively. 
Notice that the set $X$ that $G$ acts on is not required to be a group.

In this paper, we require that $\sgroup$ acts on the  setting $(\prefspace,\decspace)$, i.e., $\sgroup$ acts on both  $ \prefspace$ and $\decspace)$. For such settings, anonymity, neutrality, and resolvability are defined naturally as follows.  

\begin{dfn}[{\bf Anonymity, neutrality, and resolvability for general preferences and decisions}{}] Given the preference space $\prefspace$ and the decision space $\calD$, both of which $\sgroup$ acts on,  for any irresolute rule $\cor$ and any  profile $P$, we define 
\begin{itemize}
\item $\ano(\cor,P) \triangleq  1$ if  for any profile $P'$ with $\hist(P')=\hist(P)$, we have $\cor(P')=\cor(P)$; otherwise $\ano(\cor,P) \triangleq  0$.
\item $\neu(\cor,P) \triangleq  1$ if  for every permutation $\sigma$ over $\ma$, we have $r(\sigma(P))=\sigma(r(P))$; otherwise $\neu(\cor,P) \triangleq 0$. 
\item $\res(\cor,P) \triangleq  1$ if $|\cor(P)|=1$; otherwise $\res(\cor,P) \triangleq  0$. 
\end{itemize}
If $\ano(\cor,P) = 1$ (respectively, $\neu(\cor,P) = 1$ or $\res(\cor,P) = 1$), then we say that $\cor$ satisfies anonymity (respectively, neutrality or resolvability) at $P$. We further define $\anr(\cor,P)  \triangleq \ano(\cor,P)\times \neu(\cor,P)\times \res(\cor,P)$.  Given $n$, we say that $\cor$ satisfies anonymity (respectively, neutrality, resolvability, or ANR) if and only if for all $n$-profiles $P$, we have $\ano(\cor,P) = 1$ (respectively, $\neu(\cor,P) = 1$, $\res(\cor,P) = 1$, or $\anr(\cor,P)=1$). 
\end{dfn}

\subsection{Most Equitable Refinements For General Settings} 
\label{sec:general-MFP}
The notions of {stabilizer}, orbit, and {fixed point} in Definition~\ref{dfn:fw} can be naturally extended to the general setting $(\prefspace,\decspace)$ that $\sgroup$ acts on. For completeness, we recall the general group theoretic definitions of them below. 
\begin{dfn}[{\bf\boldmath Stabilizer, orbit, and fixed point}{}]
\label{dfn:stab}
For any group $G$ that acts on $X$, any $x\in X$, and any subset $X'\subseteq X$, 
define
\begin{align*}
\text{\bf Stabilizers of $Y$ under $G$: } &\stab_G(X') \triangleq\{g\in G: \forall x\in X', g(x)=x\}\\
\text{\bf Orbit of $x$ under $G$: } &\orbit_G(x) \triangleq\{ g(x): g\in G\}\\
\text{\bf Fixed points of $G$ in $X$: } &\fixed_G(X)\triangleq \{x\in X: \forall g\in G, g(x) = x\}
\end{align*}
The subscript $G$ is omitted when $G = \sgroup$.  
\end{dfn}

The fixed point decisions can also be defined similarly as follows. 

\begin{dfn}[{\bf\boldmath Fixed-point decisions for general settings}{}]
\label{dfn:fw} 
Given any $(\prefspace,\decspace)$ that $\sgroup$ acts on, for  any $n$-profile $P\in \prefspace^n$ and any set of decisions $D\subseteq\decspace$, define
\begin{align*} 
\text{\bf Fixed-point decisions: }&\fw(P)\triangleq  \fixed_{\stab(\hist(P))}(\calD)
\end{align*} 
\end{dfn}

Using these definitions, Lemma~\ref{lem:anr-characterization} (existence of most equitable refinements) can be extended to all $(\prefspace, \decspace)$ that $\sgroup$ acts on, formally stated in the following lemma, whose proof is similar to the proof of Lemma~\ref{lem:anr-characterization}.  
\begin{lem}[{\bf Existence of most equitable refinements, general settings}{}]
\label{lem:anr-characterization-general}
For any $(\prefspace,\decspace)$ that $\sgroup$ acts on and any anonymous and neutral rule $\cor$,   most-equitable refinements  of $\cor$ exist. Moreover,
$$\problematic{\cor}{n}{\prefspace}{\decspace} = \left\{P\in\decspace^n: \fw(P)\cap\cor(P) =\emptyset \right\}
,$$
and for every most equitable refinement $r^*$ and every $P\notin \problematic{\cor}{n}{\prefspace}{\decspace}$, $ r^*(P) \subseteq \fw(P)\cap\cor(P)$.
\end{lem}


\subsection{ANR Impossibility For General Settings} 
Following a similar proof as the proof of Lemma~\ref{ANR-Cond}, we have the following characterization of the ANR impossibility using the two constraints in  Definition~\ref{dfn:problematic-perm-group} for general settings.

\begin{thm}[{\bf\boldmath  ANR impossibility for general settings}{}]
\label{thm:ANR-general}
For any $m\ge 2$, $n\ge 1$,  and any $(\prefspace,\decspace)$ that $\sgroup$ acts on,  the ANR  impossibility holds if and only if there exists a permutation group $G\leqslant \sgroup$ such that
\begin{itemize}
\item {\bf The $\decspace$ constraint: }  $\fixed_G(\decspace)=\emptyset $.
\item {\bf The change-making constraint: }   $n$  is feasible by  $\orbsize{G}{\prefspace}$.
\end{itemize}
\end{thm}
\begin{proof}
Following a similar reasoning as that in the proof of Theorem~\ref{thm:ANR-common}, we have that the ANR impossibility holds if and only exists a permutation group $G\leqslant \sgroup$ that satisfies the $\decspace$-constraint and 
$$\text{\bf The $\prefspace$-histogram constraint: } \fixed_G(\histset{m}{n}{\prefspace})\ne\emptyset$$
The theorem follows after the following lemma, which proves the equivalence between the $\prefspace$-histotgram constraint and the change-making constraint.
\begin{lem}
\label{lem:change-making-general}
For any $(\prefspace,\decspace)$ that $\sgroup$ acts on, any $G\leqslant \sgroup$, and any $n\ge 1$,
$$\fixed_G(\histset{m}{n}{\prefspace})\ne\emptyset \Longleftrightarrow n\text{ is feasible by } {\orbsize{G}{\prefspace}}$$
\end{lem}
\begin{proof}The ``$\Rightarrow$'' direction. Let $P$ denote any $n$-profile such that $\hist(P)$ is a fixed point under $G$. Then, for any $R\in \prefspace$ and any $g\in G$, we have $[\hist(P)]_{R} = [\hist(P)]_{g(R)}$. In other words, preferences in the same orbits appear the same number of times in $P$. Let $O_1,\ldots, O_T$ denote the orbits in $\listset \ell$ under $G$ and  for every $1\le t\le T$, fix an arbitrary $\ell$-list $R^*_t\in O_t$. We have $\sum_{t=1}^T|O_t|\times [\hist(P)]_{R^*_t} = n$. Therefore,   $n$  is feasible by $\{|O_1|,\ldots, |O_T|\}$. 

The ``$\Leftarrow$'' direction. Let  $O_1,\ldots, O_T$ denote the orbits in $\listset \ell$ under $G$. Suppose $n = \sum_{t=1}^T\alpha_t\times |O_t| $, where $\alpha_1,\ldots,\alpha_T$ are non-negative integers. Let 
$$P \triangleq \bigcup\nolimits_{t=1}^T \alpha_t\times O_t$$
It follows that $\hist(P)$ is a fixed point under $G$, which proves that $\fixed_G(\histset{m}{n}{\prefspace})\ne\emptyset$.
\end{proof}
\end{proof}

As an example, we consider the setting where the preference (respectively, decision) space is the union of finitely many non-overlapping spaces, each of which $\sgroup$ acts on. That is,
$$\prefspace = \bigcup\nolimits_{j=1}^{n^*} \prefspace_{j} \text{ and } \decspace = \bigcup\nolimits_{i=1}^{m^*} \decspace_{i}$$
$\sgroup$ naturally acts on $\prefspace$ and $\decspace$ by extending its action on $\prefspace_{j}$'s and $\decspace_{i}$'s.
\begin{ex} $\listset {\le L} = \bigcup_{\ell=1}^L \listset {\ell}$.
\end{ex}

\begin{lem} 
\label{lem:ANR-union}
Let $\prefspace = \bigcup\nolimits_{j=1}^{n^*} \prefspace_{j}$ and $\decspace = \bigcup\nolimits_{i=1}^{m^*} \decspace_{i}$ be unions of disjoint sets that $\sgroup$ acts on. The  ANR impossibility holds if and only if there exists a permutation group $G\leqslant \sgroup$ such that (i) for every $i\le m^*$, $G$ has no fixed point in $\decspace_i$ and (ii) $n\in \intcomb{\bigcup_{j=1}^{n^*}\orbsize{G}{\prefspace_j}}$.
\end{lem}
\begin{proof}
The lemma follows after a straightforward application of Theorem~\ref{thm:ANR-general} by noticing that (1) $G$ has no fixed point in $\decspace$ if and only if for all $i\le m^*$, $G$ has no fixed point in $\decspace_i$, and (2) $\intcomb{\orbsize{G}{\prefspace}} = \intcomb{\bigcup_{j=1}^{n^*}\orbsize{G}{\prefspace_j}}$.
\end{proof}
That is, the ANR impossibility holds for $(\prefspace,\decspace)$ rules if and only if there exists $G$ that satisfies all constraints of the $\decspace_i$'s such that $n$ is feasible by using all coins made by $\prefspace_j$'s. Therefore, it is easier for the ANR impossibility to hold for larger $\prefspace$ and smaller $\decspace$.


\subsection{Most-Favorable-Permutation Tie-Breakings for General Settings}
\label{sec:general-MFP}
In this subsection, we show that MFP tie-breaking can be naturally extended to general settings to obtain most equal refinements as well.  Like Definition~\ref{dfn:ext-priority}, we will extend a priority order $\rhd\listset m$, w.l.o.g.~$\rhd = [1\succ 2\succ\cdots\succ m],$ to any set $X$ that $\sgroup$ acts on, especially $X\in\{\prefspace,\decspace\}$). 

First, we partition $X$ into orbits $X = O_1\cup \ldots\cup O_S$ under $\sgroup$, define an arbitrary order over orbits, e.g., $O_1\rhd\cdots\rhd O_S$, and for each orbit $s\le S$ define a ``best'' element $x_s^*$. Then, when comparing $x,x'\in X$, the element in the orbit with higher priority is more preferred; and if both are in the same orbit $O_s$, compare them w.r.t.~the distance to $x_s^*$, which is defined to be the highest-priority permutation in $\sgroup$ that maps $x$ (respectively, $x'$) to $x_s^*$. Formally, when $x$ and $x'$ are in the same orbit $O_s$, $x\rhd x'$ if and only if 
$$\arg\max\nolimits^{\rhd}_{\sigma\in \sgroup} (\sigma(x) = x_s^*) \rhd \arg\max\nolimits^{\rhd}_{\sigma\in \sgroup} (\sigma(x') = x_s^*)$$
The order over $\prefspace$ can be naturally extended to histograms as done in Section~\ref{sec:alg}. This extends $\hpp$  to $(\prefspace,\decspace)$ that $\sgroup$ acts on.
\begin{ex}
Let $m=4$, $X = \listset{\le 2} = \listset 1\cup \listset 2$. $\listset 1$ and $\listset 2$ are the two orbits under $\sgroup$. Suppose the priority over the two orbits are $\listset 1\rhd \listset 2$. Then, 
$$\{1\}\rhd\{2\}\rhd \{3\}\rhd \{4\}\rhd\{1,2\}\rhd\{1,3\}\rhd\{1,4\}\rhd\{2,3\}\rhd \{2,4\}\rhd\{3,4\}$$
Let $P= 2\times \{2\}+ \{3\}+2\times \{1,3\}+\{2,4\}$  and let $\cor$ denote the approval rule. Then, $\cor(P) = \{2,3\}$. It is not hard to verify that a $\hpp$ is $(3,2,1, 4)$, which means that $\hpp_f\ast\cor(P) = \{2\}$.
\end{ex}
Like Theorem~\ref{thm:MFP}, MFP tie-breaking computes a most equitable refinement under general settings, as stated in the following theorem.

\begin{thm}
\label{thm:MFP-general} For any $(\prefspace,\decspace)$ that $\sgroup$ acts on and any anonymous and neutral rule $\cor$, $\hpp_\tb$ is well-defined and $\breaking{\hpp_\tb}{\cor}$ is a most equitable refinement.
\end{thm}
The following brute-force algorithm computes MFP tie-breaking for general $(\prefspace,\decspace)$.
\begin{algorithm}[htp]
\caption{MFP tie-breaking for general $(\prefspace,\decspace)$.}\label{alg:MFP-general}
\begin{algorithmic}[1] 
\FOR{every $\sigma\in \sgroup$}
\STATE Compute  $\sigma(\hist(P))$. 
\IF{$\sigma(\hist(P))\rhd \vec h_{\max}$}
\STATE Let  $\vec h_{\max} = \sigma(\hist(P))$ and let $\sigma_{\max} = \sigma$.
\ENDIF
\ENDFOR 
\STATE Compute $\fw(P)\cap\cor(P)$
\RETURN  $\arg\max^\rhd _{d\in \fw(P)\cap\cor(P)} \sigma_{\max}(d)$
\end{algorithmic}
\end{algorithm}

The runtime of Algorithm~\ref{alg:MFP-general} is guaranteed in the following theorem.
\begin{thm}
\label{thm:mfp-general}
For any polynomially computable $f$ and any $(\prefspace,\decspace)$ that $\sgroup$ acts on, such that computing the outcome of permutation and comparing the priority of two elements for both $\prefspace$ and $\decspace$ take polynomial time,   Algorithm~\ref{alg:MFP-general} computes $\hpp_\tb$ in $m!\cdot\text{poly}(m,n)$ time. 
\end{thm}  
\begin{proof}
The for loop of Algorithm~\ref{alg:MFP-general} contribute to the $m!$ factor in the runtime, and the rest operations takes polynomial time. 
\end{proof}  

\end{document}